\documentclass[12pt]{iopart}

\usepackage{iopams}
\usepackage{cite}
\expandafter\let\csname equation*\endcsname\relax
\expandafter\let\csname endequation*\endcsname\relax
\usepackage[T1]{fontenc}

\usepackage{amsmath}
\usepackage{commath}
\usepackage{iopams}
\usepackage{graphicx}
\usepackage[breaklinks=true,colorlinks=true,linkcolor=blue,urlcolor=blue,citecolor=blue]{hyperref}
\usepackage{color}
\usepackage{mathrsfs}

\begin{document}

\title[Sequence-reading diffusion]{A simple model of a sequence-reading diffusion: non-self-averaging and self-averaging properties}

\author{Silvio Kalaj$^1$, Enzo Marinari$^{1,2}$, Gleb Oshanin$^{3,4}$ \& Luca Peliti$^5$}
\address{$^1$ Dipartimento di Fisica, Sapienza Universit\`{a} di Roma, P.le A. Moro 2, I-00185, Roma, Italy}
\address{$^2$ INFN, Sezione di Roma 1 and Nanotech-CNR, UOS di Roma, P.le A. Moro 2, I-00185, Roma, Italy}
\address{$^3$ Sorbonne Universit\'e, CNRS, Laboratoire de Physique Th\'eorique de la Mati\`{e}re Condens\'ee (UMR 7600), 4 Place Jussieu, 75252, Paris, Cedex 05, France}
\address{$^4$ Asia Pacific Center for Theoretical Physics, Hogil Kim Memorial Building 501 POSTECH,  37673, Pohang, South Korea}
\address{$^5$ Santa Marinella Research Institute, I-00058, Santa Marinella, Italy}

\begin{abstract}
Motivated by a question about the sensitivity of knots' diffusive motion
to the actual sequence of nucleotides placed on a given  DNA, here we study a simple model of a sequence-reading diffusion on a stretched chain 
with a frozen sequence of "letters" $A$ and $B$, having different interaction energies.  
The chain contains a single distortion - a hernia - which brings the two letters at its bottom together such that they interact.  Due to interactions with the solvent, 
the hernia performs a random hopping motion along the chain with the transition rates dependent
 on its actual position.  
Our two focal questions are a) the dependence of various transport properties on the letters' interaction energy and b) whether these  properties are self-averaging with respect to different realizations of sequences. We show that the current through a finite interval, the resistance of this interval and the splitting probabilities on this interval lack self-averaging. On the contrary, the mean first-passage time through a finite interval with $N$ sites and the diffusion coefficient in a periodic chain are self-averaging in the limit $N \to \infty$. Concurrently,  two latter properties exhibit sample-to-sample fluctuations for finite $N$, as evidenced by numerical simulations. 
\end{abstract}

Keywords:  Diffusion in disordered media, self-averaging vs non-self-averaging

\section{Introduction}

Albeit the DNA molecules are stiff at short length scales, they are typically very long and therefore become susceptible to self-entanglement and knotting \cite{lim,orl}.  Indeed, any self-avoiding chain becomes knotted with probability $1$ when its length tends to infinity \cite{sum}, which signifies that knots are inevitably present in long DNAs. Concurrently, 
knots also occur on DNAs as side-products
of elementary biological processes, such as replication, transcription and recombination (see, e.g., \cite{lim,orl}). The presence of knots is very important because, e.g., it 
reduces the mechanical strength of a chain \cite{35}, and also imposes additional constraints
on a translocation of DNAs in narrow channels \cite{rosa} and in gels \cite{10}. As well, it affects  the folding kinetics of DNAs \cite{32}  and
plays an important role in gene regulation since the knots typically 
separate topologically different parts of the genome \cite{7}.  Complexity of knots appearing on DNAs, as 
well as many ensuing 
consequences of their presence are discussed in recent reviews \cite{lim,orl}.

Whenever a DNA is not too over-stretched such that the knot is not too tightened, the latter is not localized but rather performs a long-range diffusive motion along the chain - a fascinating dynamical process which has many important implications (see above). As such, it has 
attracted a great deal of attention in the past (see, e.g., \cite{lim,orl} for a review). 
Diffusion of knots 
has been studied experimentally under different physical 
conditions (see, e.g., \cite{orl,bao,ma}) and also numerically, using  Brownian dynamics simulations (see, e.g., \cite{orl,vol,mich}). In parallel, several works focused on the theoretical understanding of mechanisms of knot's diffusion. References \cite{kar,gros} suggested a self-reptation mechanism 
according to which the diffusive dynamics of a knot originates from a snake-like motion of the chain itself. 
Another line of thought advocated in \cite{metz} proposed an alternative mechanism showing that the diffusive motion of a knot may be due to 
a knot region "breathing". However, it still remains quite a controversial issue which of the mechanisms is the dominant one, or if they act in parallel as argued in \cite{gerl}. An experimental verification may appear quite difficult because it
would require 
considering a wide range of molecular weights. In addition,  the data garnered in experiments is typically very noisy and exhibits a significant 
scatter (see, e.g.,  \cite{ma}), which does not permit to make fully conclusive statements.

There is, however, another potentially important issue which was not taken into account in the available numerical and theoretical analyses. Namely, the DNA is a hetero-polymer carrying its unique sequence of different nucleotides (letters) with different interaction energies. Because dynamics is essentially (quasi) one-dimensional, a random translocation of a knot along the chain, regardless of the actual transport mechanism, will necessarily bring different letters of the sequence in the vicinity of a knot together, affecting therefore the local transition probabilities. Since the sequence is frozen, a knot will re-read essentially the 
same part of the sequence of letters when it returns to an already visited region after 
it displaces away of it.

On intuitive grounds, one cannot expect that such sequence-reading interactions will lead to an anomalous diffusion.  Indeed, as evidenced by experimental analyses of a knot dynamics on a DNA \cite{orl,bao,ma}, the knots' mean-squared displacements $\overline{X^2(t)}$ obey the law of standard diffusion $\overline{X^2(t)} \sim t$, with $t$ being time and the bar here and henceforth denoting the average over thermal histories for a given realization of disorder.
One may rather expect that due to the presence of a random sequence of letters 
the standard characteristics (see below) of such a random motion will a) acquire a strong dependence on the interaction energies of letters and b)
will vary noticeably 
 from sample to sample, resulting potentially in a lack of self-averaging. If this is the case, one may expect to observe a very large scatter of values of the property under study, when measurements are performed on different polymers.
 
 As a matter of fact, for a related problem of the diffusive motion of a protein along a DNA
  a crucial effect of such interactions is well-appreciated. Here, the specific part (see, e.g., \cite{berg,hwa}) of the interaction energy directly associated with  the actual nucleotide sequence 
is known to produce a 
strong sequence-dependent behavior of the mean first-passage time from the initial position to a target site located some finite (but not too large) distance away of it,   
giving rise to strong sample-to-sample fluctuations \cite{slut1} (see also \cite{slut2,ralff}). 
We finally remark that an \textit{anomalous} diffusive motion due the presence of a random sequence of letters may indeed take place in somewhat more complicated physical situations, when other important factors come into play.  In particular, it is known that
 twist-induced denaturation bubbles in long DNAs exhibit aging and
move around \textit{sub-diffusively} with continuously varying dynamic
exponents \cite{hwa2}. As well, it was realized that the trajectories of the boundary between the coil and ordered helix phases in melting hetero-polymers become logarithmically confined  \cite{pgg} (see also \cite{sid}).

\begin{figure}[ht]
\centering
\includegraphics[width=30mm,angle=-90]{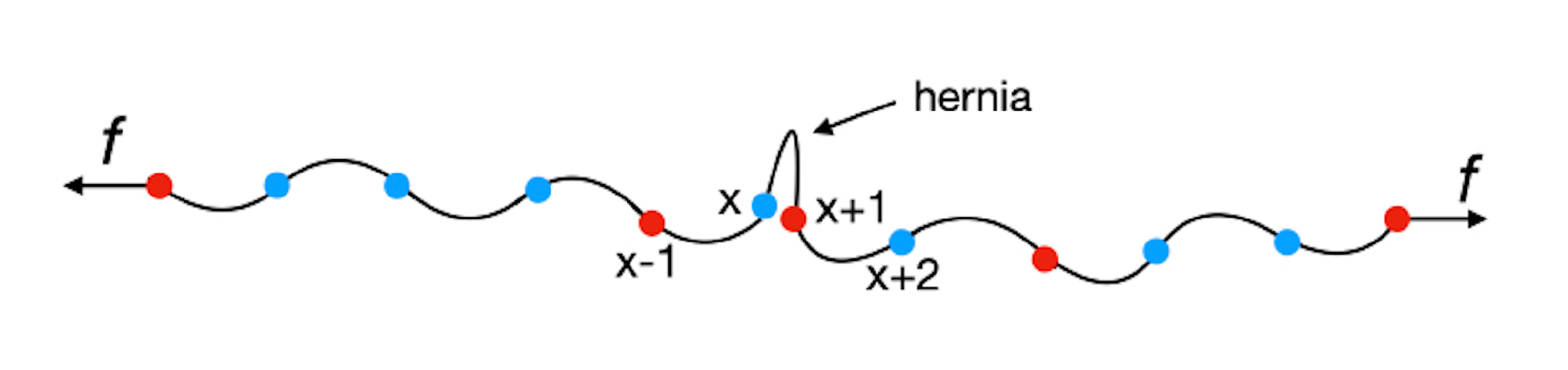}  
\caption{A stretched polymer chain with a frozen sequence of letters $A$ and $B$ (filled colored circles placed at sites $x$ along the chain, and encoded by the spin variable $\sigma_x  = \pm 1$) chosen at random, with equal probabilities $= 1/2$. The chain contains a single distortion -    
a hernia, which shortens the distance between the letters appearing at $x$ and $x+1$, such that these two letters become interacting. Other letters are placed sufficiently far from each other and are not interacting. The hernia performs an  unbiased hopping motion along the chain with the transition probabilities dependent on the energies of four letters in its immediate vicinity.}
\label{fig:1} 
\end{figure}

In quest for a possible lack of self-averaging (or conversely, self-averaging) of different sequence-dependent observables 
characterizing hernia's random motion,  in this paper we make a first step in this direction and 
introduce a minimal model of a sequence-reading diffusion. Our model is sufficiently simple to be solved exactly but  captures some important aspects and therefore permits to gain a conceptual understanding of the effect of a random sequence of letters on the dynamics of knots. More specifically, 
we consider a finite-size 
chain immersed in a heat bath and stretched by two forces applied to both extremities of the chain and pointing in the opposite directions (see Fig. \ref{fig:1}). The chain bears a quenched sequence of letters $A$ and $B$, placed at random and independently of each other, present in the same amounts, on average, and having different interaction energies. The chain contains a single distortion - not a real knot but rather a hernia - in which the two letters at its bottom appear sufficiently close to each other and interact,  while other letters within the sequence are assumed to be placed sufficiently far from each other and hence, to be non-interacting. The hernia's random motion is supposed to be an unbiased, thermally-activated process  with the transition probabilities dependent on the actual position within the sequence, as will be specified below. Lastly, we remark that if we were to think about an experimental realization of such a model, we would have to take into account that the hernia can spontaneously disappear and re-appear on the chain at some other place. However, being motivated by the problem of a sequence-reading dynamics of knots, which cannot  disappear spontaneously, we assume that the life-time of a hernia is infinitely long and therefore it cannot disappear - just to translate randomly along the chain.

To quantify the effect of a random sequence of letters on the transport properties, we introduce the parameter 
\begin{align}
\label{R}
R_{\zeta} = \frac{\langle \zeta^2 \rangle - \langle \zeta \rangle^2}{\langle \zeta \rangle^2}  \,,
\end{align} 
where the angle brackets here and henceforth denote averaging with respect to different realizations of disorder, while  $\zeta$ is a random (sequence-dependent) variable of interest. This parameter was standardly used (see, e.g., \cite{pastur,domany}) for the analysis of $d$-dimensional disordered systems to distinguish between three possible scenarios: the ratio $R_{\zeta}$ (here $\zeta$ can be, e.g., internal energy, magnetization, specific heat or susceptibility, if one deals with a disordered system with $N$ interacting spins)  
approaches a constant value in the limit $N \to \infty$, in which case $\zeta$ is \textit{not self-averaging}; 
$R_{\zeta} \sim 1/N^{\alpha}$ with $0 < \alpha < d$ in which case $\zeta $ is \textit{weakly self-averaging} and eventually, $\zeta$ is \textit{strongly
 self-averaging} when $R_{\zeta} \sim 1/N^d$. 
 Clearly, in a self-averaging system it is enough to have a single, sufficiently large sample and the behavior observed on this sample will nonetheless be representative of a statistical ensemble of all possible samples. In the absence of self-averaging  the situation is completely different; here, on the contrary, the value of an observable deduced from 
 measurements performed on a single sample, arbitrarily large, will not give 
 a meaningful result and must be repeated on an ensemble of samples.

Random \textit{sequence-dependent} variables on which we concentrate here are : \\
a)  Resistance $\tau_N$  of a finite chain of length $N$ with respect to a random transport. \\
b)  Its reciprocal value - the current $j_N$ through a finite chain of length $N$. \\
c)  Splitting probability $E_-$, i.e.,  the probability that for a diffusion on a finite chain the left end of the latter will be reached first without ever hitting its right end. \\
d) Mean\footnote{The term "mean" here and henceforth corresponds to averaging over thermal noises.} first-passage time $T_N$ through a finite chain with $N$ sites. \\
e)  Diffusion coefficient $D_N$ in a periodic chain with $N$ sites.  

We proceed to show in what follows that the first three random variables; namely, 
 the resistance, the current through a finite chain and the splitting probability are \textit{not self-averaging} in the limit $N \to \infty$, 
while the remaining two, i.e.,  the mean first-passage time and the diffusion coefficient are strongly self-averaging in the limit $N \to \infty$. This signifies that the three former properties will exhibit strong sample-to-sample fluctuations for any value of $N$, while the two latter are, of course, fluctuating for the intermediate values of $N$ (and these sample-to-sample fluctuations can be quite significant, as will be evidenced via numerical simulations) but the effect of fluctuations will become insignificant in the  limit $N \to \infty$.  We note that while it is absolutely legitimate to consider the behavior of $D_N$ in the limit $N \to \infty$,  because it permits to determine the effective diffusion coefficient in an infinite system, this is evidently not so when we analyze the mean first-passage time. Indeed, for the latter $N$ is essentially finite in most of applications and the knowledge of the behavior in the limit $N \to \infty$ is therefore only of a conceptual importance, pointing on a general trend.
 
The paper is outlined as follows: In Sec. \ref{sec:2} we describe our model in more detail. In Sec. \ref{sec:3}
we discuss the behavior of the resistance, the probability current and the splitting probability, determine the explicit dependences of their first two moments on the interaction parameters and show that these properties are \textit{not self-averaging}. We also evaluate numerically the probability density function of the resistance to show that this function is three-modal, which explains why the resistance (as well as two other above properties which are mathematically linked to each other) lacks self-averaging. 
In Secs. \ref{sec:4} and \ref{sec:5} we focus on the sequence-dependent mean first-passage time through a finite interval and the  sequence-dependent diffusion coefficient in a periodic chain, respectively. We determine the dependence of the moments of these characteristic properties on the interaction energy and demonstrate that they are self-averaging in the limit $N \to \infty$. 
Behavior at intermediate length scales is analyzed via numerical simulations which reveal significant sample-to-sample fluctuations. 
We conclude in Sec. \ref{sec:6} with a brief recapitulation of our results and a discussion.
In the main text we merely explain the steps involved in the derivation of our final results, present these results and their analysis; details of the derivations are relegated to Appendices.

\section{Model}
\label{sec:2}

Consider a single polymer chain of a finite extent immersed in a solvent, which acts as a heat bath (see Fig. \ref{fig:1}). The
chain is stretched by external forces $f$ acting on its end-points (which can be experimentally realized, e.g., by attaching 
beads to the end-points of the chain and manipulating with them by optical tweezers) and are pointing in the opposite directions. 
In such a situation, the chain can be considered as a one-dimensional lattice of sites, appearing at some distance $a$ from each other and labelled along the chain by the integer variable $x$. Without any lack of generality, we set $a = 1$. 

Suppose that, as it happens in the above-mentioned situation with a DNA, 
the chain is not chemically-homogeneous but carries a random quenched sequence of  different chemical groups. 
Here we will focus on the simplest case when there are only two such chemical groups, which we label by   
letters $A$ and $B$. 
We assume then that at each site $x$ there is   
either of the letters $A$ or $B$, chosen at random with equal probability $= 1/2$. To take into account the presence of
such a heterogeneity, we 
 assign to each site $x$ a quenched, random two-state variable $\sigma_x$ :  
 \begin{equation}
 \sigma_x = \begin{cases}
\, \, 1  & \,\,\,\, \text{letter $A$, \, prob $1/2$} \,, \\
-1 & \,\,\,\, \text{letter $B$, \,  prob $1/2$} \,.
\end{cases}
\end{equation}

Assume next that there is a single perturbation of the chain - a hernia,  which effectively shortens the spatial distance between the two letters placed at $x$ and $x+1$ bringing them in contact (see Fig. \ref{fig:1}). In this way,  these two letters start to "feel" each other and
therefore interact, as opposed to other letters which are not sufficiently close to each other and hence, are not interacting. In what follows, we adopt a convention that the position of the hernia is defined by the site from its left, i.e., for the situation depicted in Fig. \ref{fig:1} the hernia is located at the site $x$.

Denoting 
the interaction energy of an $AB$ pair as $u_{AB}$, and assuming, for simplicity, that the interaction energies of the pairs $AA$ and $BB$ are equal to the same value $u_0$, 
we represent the energy $U_x$ of the hernia being at site $x$ as
\begin{align}
\label{1}
U_{x} = J \, \sigma_x \sigma_{x+1} + K  \,,
\end{align}
where the "Ising" coupling constant $J$ and the constant $K$ are given by
\begin{align}
\label{00}
J = \frac{u_0 - u_{AB}}{2} \,,  \quad K  =  \frac{u_0 + u_{AB}}{2}  \,.
\end{align}
To somewhat ease the notations, in what follows we will also use the dimensionless parameters
\begin{align}
\label{000}
J' = \frac{\beta J}{2} \,,  \quad K'  =  \frac{\beta  K}{2}  \,,
\end{align}
where $\beta = 1/k_B T$ denotes the reciprocal temperature measured in units of the Boltzmann constant $k_B$.
\begin{figure}[ht]
\centering
\includegraphics[width=150mm]{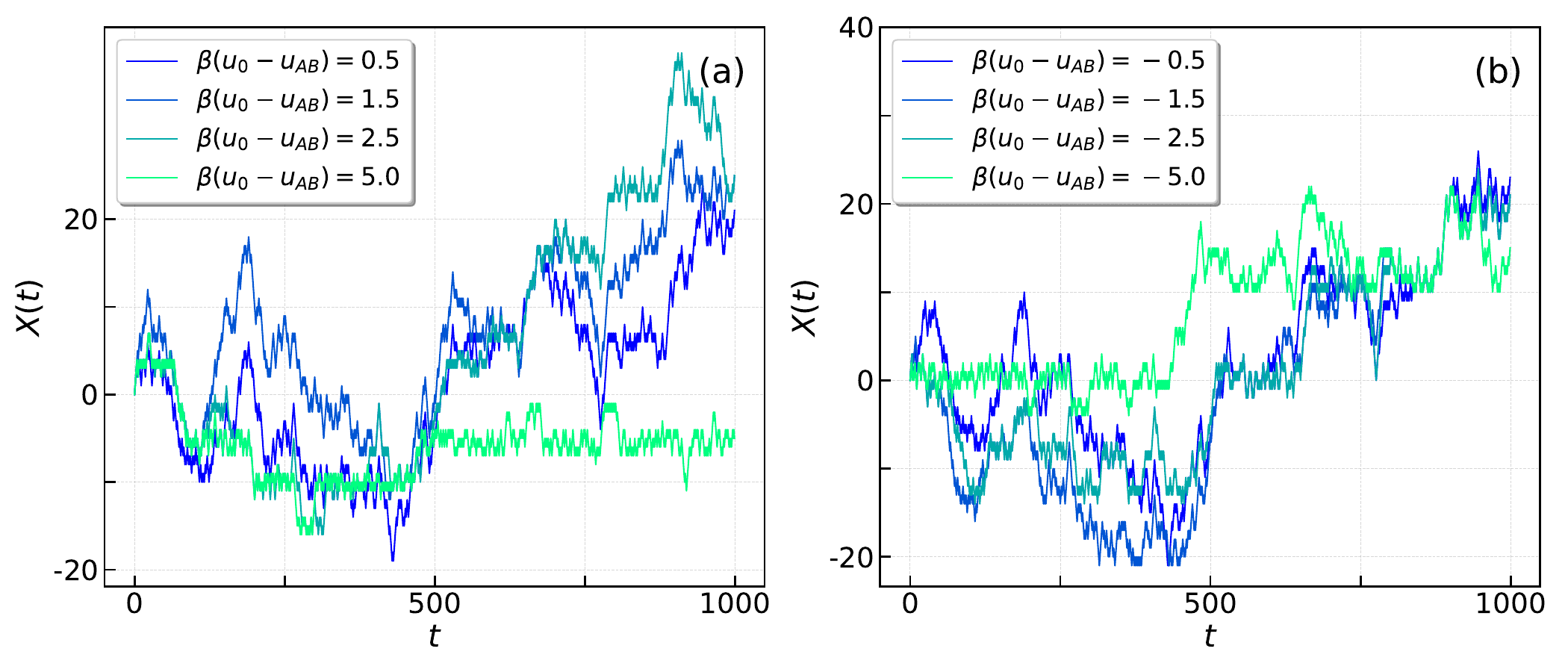}  
\caption{Trajectories $X(t)$ of the hernia starting at the origin at time $t=0$ 
on a chain with a given, randomly-generated sequence of letters $A$ and $B$ for several 
values of parameter $\beta J$. 
Panel (a):  $\beta J > 0$ ($u_0 >  u_{AB}$). Panel (b): $\beta J  < 0$  ($u_0 < u_{AB}$).  
 }
\label{fig:2} 
\end{figure}

Lastly, suppose that due to interactions with the heat bath the hernia is not localized, but performs an unbiased, thermally-activated hopping motion along the chain in a continuous time.  Then, the time evolution of the probability $P_x = P(x,t)$
of finding the hernia on site $x$ at time instant $t$ is governed by the Master equation of the form
\begin{align}
\label{ME}
\dot{P}_x = - \left(\alpha_x + \beta_x\right) P_x + \alpha_{x-1} P_{x-1} + \beta_{x+1} P_{x+1} \,, 
\end{align}
where the dot denotes the time derivative, while $\alpha_x$ and $\beta_x$ are
 the forward and backward, respectively, hopping rates from the site $x$. These rates are equal to the probabilities $p_{x,x+1}$
and $p_{x,x -1}$, respectively, 
of jumps from $x$ to $x+1$ and to $x-1$ per unit time $\delta t$:
\begin{align}
\label{rates}
\alpha_x = \frac{p_{x,x+1}}{\delta t} \,, \quad \beta_x = \frac{p_{x,x -1}}{\delta t}  \,.
\end{align}
The transition probabilities $p_{x,x+1}$  and $p_{x,x-1}$ are defined, in the standard way, as
\begin{equation}
\begin{split}
\label{3}
p_{x,x+1} &= Z^{-1}_x   \exp\left(\frac{\beta}{2}\Big[U_{x} - U_{x+1}  \Big]\right) \\&=Z^{-1}_x \exp\Big(J' \sigma_{x+1} \left(\sigma_{x} - \sigma_{x+2}\right) \Big) \,, \\
p_{x,x-1} &= Z^{-1}_x \exp\left(\frac{\beta}{2} \Big[U_{x} - U_{x-1} \Big] \right) \\
&= Z_x^{-1} \exp\Big(J' \sigma_{x} \left(\sigma_{x+1} - \sigma_{x-1}\right) \Big) \,, 
\end{split}
\end{equation}
 where $Z_x$ is the normalization which ensures that
 $p_{x,x+1} + p_{x,x-1} \equiv 1$, (i.e., the hernia is not pausing at $x$), so that
 \begin{align}
 Z_x = \exp\Big(J' \sigma_{x+1} \left(\sigma_{x} -  \sigma_{x+2}\right)\Big) + 
 \exp\Big(J' \sigma_{x} \left(\sigma_{x+1} - \sigma_{x-1}\right)\Big) \,.
 \end{align}
Note that according to the above definition, 
the hernia has a tendency to jump from $x$ towards the site on which it will have a lower energy, if $U_{x-1} \neq U_{x+1}$. If $U_{x-1} = U_{x+1}$, one has that $p_{x,x+1} = p_{x,x-1} = 1/2$, such that the hernia chooses the jump direction with probability $=1/2$.

Eventually, noticing that some of the terms in eqs. \eqref{3} cancel each other, we have that $p_{x,x+1} $ and $p_{x,x-1}$ are given explicitly by
\begin{align}
\begin{split}
\label{33}
p_{x,x+1} &= \frac{\exp\Big(-J' \sigma_{x+1} \sigma_{x+2}\Big)}{\exp\Big(-J' \sigma_{x+1} \sigma_{x+2}\Big)+ \exp\Big(-J' \sigma_{x-1} \sigma_{x}\Big)} = \frac{\phi_{x-1}}{\phi_{x-1} + \phi_{x+1}}  \,,\\
p_{x,x-1} &= \frac{\exp\Big(-J' \sigma_{x-1} \sigma_{x}\Big)}{\exp\Big(-J' \sigma_{x+1} \sigma_{x+2}\Big)+ \exp\Big(-J' \sigma_{x-1} \sigma_{x}\Big)} = \frac{\phi_{x+1}}{\phi_{x-1} + \phi_{x+1}} \,,
\end{split}
\end{align}
where we have used the shortened notation
\begin{align}
\label{phi}
\phi_x = \exp\Big(J' \sigma_x \sigma_{x+1}\Big) \,.
\end{align}
In what follows, we will  first seek convenient exact representations of the realization-dependent random variables under study  in terms of variables $\phi_x$, and then will perform averaging over the  "spin" variables $\sigma_x$. 

We conclude this section with the following remarks : the dynamics of the hernia in our model is  
an unbiased hopping motion in presence of a \textit{random} potential which is a combination of finite-depth traps and finite-height barriers. The transition probabilities from the site $x$ are not local, i.e., dependent only on the letter which appears at the site $x$, but 
also depend explicitly on the letters which are placed on the sites $x-1$, $x+1$ and $x+2$. This implies that the random potential in which the hernia moves is \textit{correlated}, with correlations extending over a finite interval. 
To gain some preliminary insight, we performed numerical simulations of the hernia's hopping motion 
(with the hopping probabilities defined in eqs. \eqref{33}) on a chain with a given, randomly-generated sequence of letters,   
for several positive and negative values of the parameter $\beta J$. Results of these simulations are presented in 
Fig. \ref{fig:2} in which we depict 
individual (for a given thermal history) "short" trajectories $X(t)$ of the hernia. 
 We observe that for sufficiently low values of 
$\beta J$, both positive and negative, the trajectories $X(t)$ become 
delocalized at relatively short times and behave essentially like standard random walk trajectories. 
For $\beta J = \pm 2.5$ the trajectories are localized for extended (random) periods of time between positions of the $AB$ pairs but ultimately get delocalized, as they should. On contrary, for $\beta J = \pm 5$ the trajectories are localized over the entire observation time-interval. Here, one has to go to much larger time-scales in order to observe a 
crossover to the
 standard random walk behavior. 
 This signifies that the typical jump-time, (and hence, the diffusion coefficient) becomes strongly dependent on the interaction parameter $J$.

\section{The sequence-dependent resistance, the probability current and splitting probabilities in a finite chain} 
\label{sec:3}

Consider a chain in which we keep a fixed value of the probability $P_{x=0} = P(x=0,t) = c_0$ at 
the site $x=0$, and place a perfect trap  
at the  site $x = N$,  at which the hernia is removed from the chain; that being, $P_{x=N} = P(x=N,t) = 0$. In such a system, the sequence-dependent probability current $j_N$ through the interval $(0,N)$  attains in the steady-state a well-known form (for completeness, we present in \ref{A} details of the derivation, see also \cite{burl,gaveau}) 
\begin{align}
\label{j}
j_N = \frac{c_0}{2 \delta t \, \tau_N} \,,
\end{align}
where $\tau_N$ - a sequence-dependent \textit{resistance} of a finite chain with 
respect to the transport of hernias - 
 is the following
functional of the transition probabilities
\begin{align}
\label{tau}
&\tau_N = 1 + \frac{p_{1,0}}{p_{1,2}} +  \frac{p_{1,0} p_{2,1}}{p_{1,2} p_{2,3}} + \frac{p_{1,0} p_{2,1} p_{3,2}}{p_{1,2} p_{2,3} p_{3,4}} + \frac{p_{1,0} p_{2,1} p_{3,2} p_{4,3}}{p_{1,2} p_{2,3} p_{3,4} p_{4,5}} + \nonumber\\
& \ldots + \frac{p_{2,1} p_{3,2} p_{4,3} \ldots p_{N-2,N-3}}{p_{1,2} p_{2,3} p_{3,4} p_{4,5} \ldots p_{N-2,N-1 }} + \frac{p_{2,1} p_{3,2} p_{4,3} \ldots p_{N-2,N-3} p_{N-1,N-2}}{p_{1,2} p_{2,3} p_{3,4} p_{4,5} \ldots p_{N-2,N-1} p_{N-1,N}}  \,.
\end{align}
Below we study positive moments of the realization-dependent resistance $\tau_N$ 
and its negative moments, which are the moments of the realization-dependent steady-state current $j_N$. As well, we will consider 
the moments of the splitting probabilities, which are functionals of $\tau_N$ (see below).

\subsection{Sequence-dependent resistance $\tau_N$.}
Random variables of the form as in eq. \eqref{tau}, which are known as Kesten variables \cite{3}, emerge in different contexts in physics. 
Their properties has been analyzed in case when 
the ratios $p_{x,x-1}/p_{x,x+1}$ are independent, identically distributed random variables 
(see, e.g., \cite{calan,flux,alain}), except for reference \cite{albe} which focused on the case 
when these ratios have strong power-law correlations. Their continuous-space counterparts were also studied in the mathematical finance literature, because they are related to some path-dependent (namely, Asian) options (see, e.g., \cite{yor,greg}).

In the situation under study the ratios $p_{x,x-1}/p_{x,x+1}$ are not independent but are sequentially correlated. To make this statement more explicit, we rewrite the expression \eqref{tau} in an equivalent form, taking advantage of the definition of the transition probabilities in eqs. \eqref{33} and \eqref{phi}. This gives
\begin{align}
\label{taup}
\begin{split}
\tau_N &= 1 + \frac{\phi_2}{\phi_0} +  \frac{\phi_2 \phi_3}{\phi_0 \phi_1} +   \frac{\phi_2 \phi_3 \phi_4}{\phi_0 \phi_1 \phi_2} + \ldots +  \frac{\phi_2 \phi_3 \phi_4 \ldots \phi_{N-1}}{\phi_0 \phi_1 \phi_2 \ldots \phi_{N-3}} +  \frac{\phi_2 \phi_3 \phi_4 \ldots \phi_{N-1} \phi_N}{\phi_0 \phi_1 \phi_2 \ldots \phi_{N-3} \phi_{N-2}}  \\
&= 1 + \frac{1}{\phi_0 \phi_1} \left(\phi_1 \phi_2 +  \phi_2 \phi_3 + \phi_3 \phi_4 + \ldots + \phi_{N-2} \phi_{N-1} + \phi_{N - 1} \phi_N   \right) \,.
\end{split}
\end{align}
Note that each term in the brackets, say $\phi_2 \phi_3$, does not decouple because $\phi_2$ and $\phi_3$ share a common spin variable $\sigma_3$, and moreover is statistically-dependent on the two neighboring terms, $\phi_1 \phi_2$ and $\phi_3 \phi_4$.

\begin{figure}[ht]
\centering
\includegraphics[width=150mm]{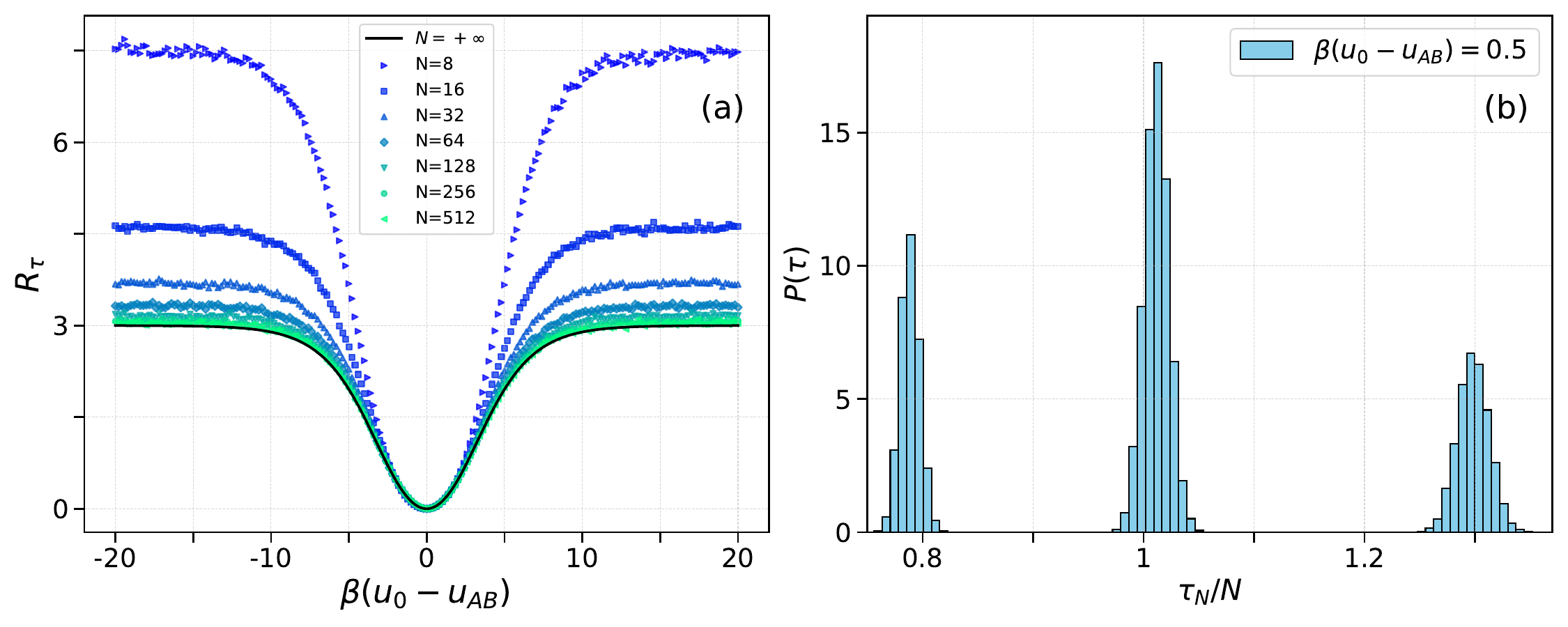}  
\caption{The resistance $\tau_N$ of a finite chain with a binary alphabet. Panel (a): The relative variance $R_{\tau}$ as function of $\beta (u_0 - u_{AB})$. Solid black curve in the bottom 
depicts  the asymptotic ($N \to \infty$) result in 
eq. \eqref{Rtau}.  Symbols present the finite-$N$ (see the inset) numerical simulations results for $R_{\tau}$  with the averaging performed over $10^5$ samples. Panel (b): The numerically-evaluated probability density function $P(\tau)$ of the rescaled resistance $\tau = \tau_N/N$ for $\beta (u_0 - u_{AB}) = 0.5$ and $N = 512$. }
\label{fig:3} 
\end{figure}

First two moments of $\tau_N$  can be calculated 
very directly by averaging the expression \eqref{taup} and its squared value. This
gives, in the limit $N \gg 1$, 
\begin{align}
\begin{split}
\label{tau12}
\Big \langle \tau_N \Big \rangle &= N \left \langle \frac{1}{\phi_0 \phi_1} \right \rangle \left \langle \phi_0 \phi_1 \right \rangle  = N \cosh^4(J')  + O\left(1\right)\,, \\
\Big \langle \tau_N^2 \Big \rangle &= N^2 \left \langle \frac{1}{\phi_0^2 \phi_1^2} \right \rangle \left \langle \phi_0 \phi_1 \right \rangle^2 = N^2 \cosh^2(2 J') \cosh^4(J') + O\left(N\right)  \,, 
\end{split}
\end{align}
where the symbols $O\left(1\right)$ and $O\left(N\right)$ signify that the omitted terms are independent of $N$ and are linear with $N$, respectively. Equations \eqref{tau12} 
yield the following expression for the relative variance (see eq. \eqref{R}), 
\begin{align}
\label{Rtau}
R_{\tau} = \frac{\cosh^2(2 J')}{\cosh^4(J')} - 1 + O\left(1/N\right)\,.
\end{align}
Therefore, the resistance $\tau_N$ is \textit{not self-averaging}, except for the trivial case when the letters $A$ and $B$ have the same interaction energy, i.e., $J'=0$.  In Fig. \ref{fig:3}(a) 
we depict the limiting form in eq. \eqref{Rtau} as function of 
$\beta (u_0 - u_{AB})$ (see the black solid curve) together with the function $R_{\tau}$ 
obtained by numerically averaging $\tau_N$ and $\tau_N^2$ for finite values of $N$. 
 We observe that, not counter-intuitively, $R_{\tau}$ is larger for smaller $N$, and converges from above to the limiting curve in eq. \eqref{Rtau} upon an increase of $N$. The limiting value of the relative variance is a symmetric function of $\beta (u_0 - u_{AB})$ and 
tends to the asymptotic value $3$ when $\beta (u_0 - u_{AB}) \to \pm \infty$, i.e., the asymptotic value of the variance is three times bigger than the squared first moment. For finite $N$ the asymptotic values of the relative variance attain even higher values.

Further on, in order to get an idea  why $\tau_N$ is not self-averaging, 
we evaluate numerically the full probability density function $P(\tau)$ of the rescaled resistance $\tau = \tau_N/N$  for 
$\beta (u_0 - u_{AB}) = 0.5$ and $N = 512$. 
The result is presented in Fig. \ref{fig:3}(b) and 
we realize that, interestingly enough,  $P(\tau)$ is not a uni-modal but a three-modal function with 
three well-separated peaks centered around the values $\tau = \tau^{(1)} = \exp( - 2 J') \cosh^2(J')$, $\tau^{(2)} = \cosh^2(J')$ and $\tau^{(3)} = \exp(2 J') \cosh^2(J')$ (see \ref{C} for the analysis of the moment-generating function of the resistance $\tau_N$). The three-modality of the probability density function is precisely the reason why this random variable is not self-averaging, even if the thickness of each of the peaks vanishes in the limit $N \to \infty$. More detailed analysis of $P(\tau)$ will be presented elsewhere
\cite{enzo4}.

 We close this subsection with a natural question:  Do fluctuations of the resistance become more pronounced in case of alphabets containing more than two letters, (which is physically a more realistic situation), or not? For the variable $\tau_N$ it is easy to find the answer  by considering, e.g., a ternary alphabet containing
$A$ and $B$ letters present at equal concentrations $=(1-p)/2$, and letters $\varnothing$, present at concentration $p$. Here, for simplicity, we suppose that $\varnothing$s are "voids", which are non-interacting neither between themselves nor with $A$s or $B$s. As above, we assume that the energy of an $AB$ pair is equal to $u_{AB}$, while the interactions energies of the $AA$ and $BB$ pairs are the same and are equal to $u_0$.
Naturally, such a situation can be also regarded as a 
chain with a randomly-rarified binary sequence. More systematic analysis of the dynamical behavior for alphabets with more than two letters will be discussed in \cite{enzo4}.

  \begin{figure}[ht]
\centering
\includegraphics[width=150mm]{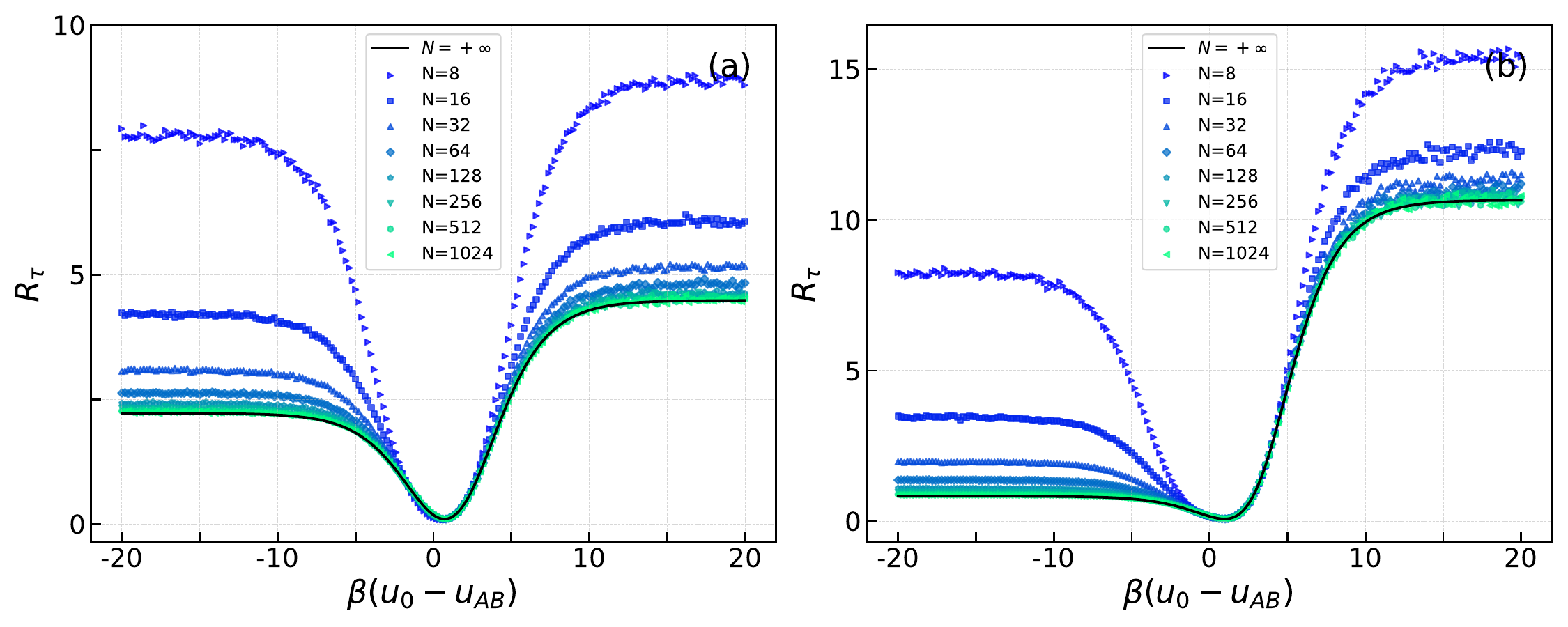}  
\caption{Relative variance of $\tau_N$ for a chain with a randomly-rarified binary sequence. Panel (a): The relative variance $R_{\tau}$ as function of $\beta (u_0 - u_{AB})$ for the concentration $p$ of voids $p=0.1$. Solid black curve in the bottom 
depicts  the asymptotic ($N \to \infty$) result in 
eq. \eqref{Rtaup}.  Symbols present the finite-$N$ (see the inset) numerical simulations results for $R_{\tau}$  with averaging performed over $10^5$ samples. Panel (b): The same for the concentration of voids $p=0.3$. Note that for a randomly-rarified binary sequence the relative valiance $R_{\tau}$ becomes an asymmetric function of $\beta (u_0 - u_{AB})$.}
\label{fig:4} 
\end{figure}
Relegating the details of calculations to \ref{B}, we have that in this case the relative variance $R_{\tau}$ is given explicitly by
\begin{align}
\label{Rtaup}
R_{\tau} = \frac{p + (1-p) \left[p + (1-p) e^{-2 K'} \cosh(2 J')\right]^2}{\left(p + (1-p) \left[p + (1-p) e^{-K'} \cosh(J')\right]^2\right)^2} - 1 + O\left(1/N\right)\,,
\end{align}
 where $K'$ is defined in eqs. \eqref{00} and \eqref{000}.
 Expression \eqref{Rtaup} reduces to the result in eq. \eqref{Rtau} when $p=0$, as it should. When $p=1$, i.e., when the system contains only voids, $R_{\tau} \equiv 0$, i.e., $\tau_N$ is not fluctuating.
We depict the above expression in Fig. \ref{fig:4}  for two values of the concentration $p$ and fixed $\beta u_0 = 1$, together with the results of numerical simulations performed for finite values of $N$. We observe that the presence of already a third letter, even when it is just a void, complicates the behavior:  $R_{\tau}$ is no longer a symmetric curve around $J' = 0$;  fluctuations become more pronounced for positive $J'$ and  are relatively less important for negative $J'$; $R_{\tau}$ is always greater than zero, even for $J'=0$, meaning that $\tau_N$ is not self-averaging for any value of the system's parameters (except for a singular case when $p\equiv1$). As in the case of a binary alphabet, for a finite concentration of voids fluctuations are more important for finite $N$ than in the limit $N \to \infty$.
To conclude, one may expect that the number of different chemical groups on a chain is indeed an important and relevant factor. Another interesting aspect may be the presence of correlations in the placement of letters \cite{madden}, which will be discussed elsewhere \cite{enzo4}.
 
\subsection{Sequence-dependent steady-state current $j_N$} 
 
 We turn next to the analysis of \textit{negative} moments of $\tau_N$, i.e., the \textit{positive} moments of the realization-dependent steady-state current in eq. \eqref{j}. This is, evidently, a much more complicated problem than the calculation of $\langle \tau_N^n\rangle$ with positive $n$, because it needs the knowledge of the full moment-generating function of $\tau_N$, 
 \begin{align}
 \label{Phi1}
 \Phi_N(\lambda) = \Big \langle \exp\left(  - \lambda \tau_N \right) \Big \rangle \,, \quad \lambda \geq 0 \,.
 \end{align}
The function defined in the above expression is calculated in \ref{C}. The moments of the current are then obtained by integration of $\Phi_N$, multiplied by $\lambda^{n-1}$ with arbitrary positive $n$, over $\lambda$ to give the following expressions for the first and second moments of the realization-dependent probability current $j_{N}$:
\begin{align}
\begin{split}
\label{momentsj}
\Big \langle j_{N} \Big \rangle &=  \frac{c_0}{2 \delta t \, N} \Bigg(1 + \frac{3 \cosh^4(J') - 4 \cosh^2(J') + 1 }{\cosh^4(J') N}
+ O\left(1/N^2\right) \Bigg) \,,\\ 
\Big \langle j_{N}^{2} \Big \rangle &=  \left(\frac{c_0}{2 \delta t \, N}\right)^2 \Bigg(\frac{\cosh^2(2 J')}{\cosh^{4}(J')} +  \frac{C_2(J')}{N} + O\left(1/N^2\right) \Bigg) \,, \\
C_2(J') & =\frac{\tanh^2(J') \cosh(2 J') \left(11 \cosh(4 J') + 2 \cosh(2 J') + 3\right)}{4 \cosh^6(J')}  \,.
\end{split}
\end{align}
Note that  the coefficients in front of the inverse powers of $N$ in the parentheses become identically equal to zero when $J'$ is set equal to zero, while the leading $N$-independent term becomes equal to $1$, as it should. 

From eqs. \eqref{momentsj} we find that the relative variance $R_j$  obeys
\begin{align}
R_j = \frac{\left \langle j_{N}^{2} \right \rangle - \left \langle j_{N} \right \rangle^2}{\left \langle j_{N} \right \rangle^2} = \frac{\cosh^2(2 J')}{\cosh^{4}(J')} - 1 + O\left(1/N\right) \,,
\end{align} 
which signifies that (likewise the resistance $\tau_N$)  the sequence-dependent stationary
 current $j_N$ is \textit{not self-averaging}. Note that the right-hand-side of the above expression 
is exactly the same function of $J'$  as $R_{\tau}$ in eq. \eqref{Rtau}.

 We notice next that the disorder-averaged current, i.e., $\langle j_N\rangle$, exhibits a rather astonishing behavior. Namely, in contrast to $\langle \tau_N \rangle$ which depends on $J'$ already in the leading in $N$ order, the disorder-averaged stationary current through a finite (albeit long) chain appears to be independent of $J'$ in this order. Indeed, the result in the first line in eq. \eqref{momentsj} shows that regardless of the actual value of $J'$, 
\begin{align}
\label{fick}
\Big \langle j_{N} \Big \rangle =  \frac{c_0}{2 \delta t \, N}  + O\left(1/N^2\right) \,,
\end{align} 
as $N \to \infty$, which is the standard Fickian result for the current in a homogeneous chain \cite{sidgen}. In other words, for $N \to \infty$ the disorder-averaged current does not "feel" any disorder. A seemingly plausible explanation of such a behavior is as follows: the first moment of the sequence-dependent current is supported by \textit{anomalous} realizations of random sequences in which the letters $A$ and $B$ are strongly clustered into islands containing either of them; in such realizations the transition probabilities within the islands are energy-independent, and hence,  symmetric, which means that within each island the hernia performs a standard random walk. The hernia changes its energy 
only at the boundaries between two neighboring islands which events contribute to the first moment of the current only 
in the sub-leading in $N$ order. In turn, the second moment of $j_N$ depends on $J'$ already in the leading order, which ensures that $j_N$ lacks self-averaging. 

\begin{figure}[ht]
\centering
\includegraphics[width=150mm]{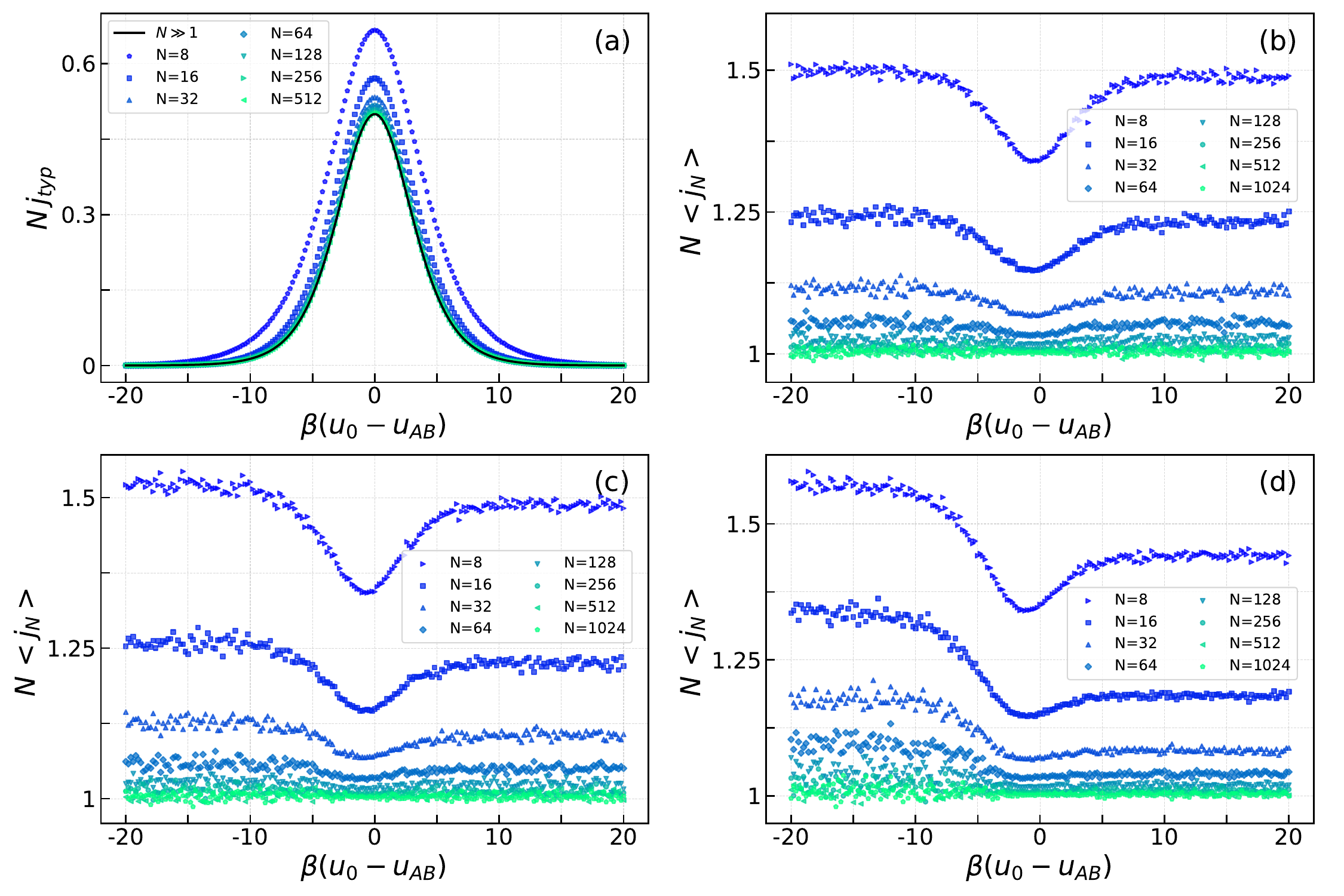}  
\caption{Steady-state probability current through a finite interval as function of $\beta (u_0 - u_{AB}) 
$ for $c_0 = 1$ and $\delta t = 1$. Panel (a): Typical current $j_{typ}$ in eq. \eqref{jtyp} multiplied by $N$ in a chain with a binary alphabet 
for different values of $N$. Black solid curve depicts the analytical result in eq. \eqref{jtyp2}. Symbols present 
$N j_{typ}$ for different values of $N$ (see the inset), calculated 
by exponentiating the numerically-averaged logarithm of $j_N$. Averaging in this panel and also in panels (b), (c) and (d) is performed over $10^5$ samples.
Panel (b): Results of numerical simulations for the first moment of the current (multiplied by $N$) in a chain with a randomly-rarified binary alphabet for a small concentration of voids, $p = 0.05$. Panels (c) and (d) present 
the first moment of the current for concentrations of voids $p=0.1$ and $p=0.3$, respectively. Note that for finite $N$ 
the averaged current $\langle j_N\rangle$ exhibits a noticeable variation with $J'$, but progressively 
looses this dependence upon a gradual increase of $N$.  
 }
\label{fig:5} 
\end{figure}

In this regard, it seems interesting to consider the "typical" current $j_{typ}$ which should be observed for most of realizations of disorder,
\begin{align}
\label{jtyp}
j_{typ} = \exp\left(\Big \langle \ln(j_N) \Big \rangle \right) \,.
\end{align}
Performing the integral in eq. \eqref{B30} (see \ref{C}) in the limit $N \to \infty$, which  defines 
 the asymptotic behavior of the moments of $j_N$ of arbitrary, not necessarily integer positive order $n$, we arrive at the following general result :
\begin{align}
\begin{split}
\label{jn}
&\Big \langle j_{N}^{n} \Big \rangle =  \left(\frac{c_0}{2 \delta t \, N}\right)^n \Bigg(\frac{\cosh^2(n J')}{\cosh^{2 n}(J')} + \frac{C_n(J')}{N}
+ O\left(1/N^2\right)\Bigg) \,, \\
C_n(J') &= \frac{n \cosh(n J') \sinh(J')}{16 \cosh^{2(2+n)}(J')} \Big((5n+1) \sinh((n+3) J') \\&- (n + 13) \sinh((n-1) J') - (5 n+9) \sinh((n-3) J') \\&+(n-11) \sinh((n+1) J')
\Big) \,.
\end{split}
\end{align}
Note that in the above expressions the coefficient $C_n(J')$ vanishes when $J' = 0$, (as well as the coefficients in all the terms proportional to higher-order inverse powers of $N$), while the first, $N$-independent term in the parentheses becomes equal to unity. 

Next, employing the usual replica-trick relation, we find
\begin{align}
\begin{split}
\label{rt}
&\Big \langle \ln(j_N) \Big \rangle = \lim_{n\to 0} \frac{\left \langle j_N^n \right \rangle - 1}{n} \\&= \ln\left(\frac{c_0}{2 \delta t \,  \cosh^2(J') N}\right) + \frac{\left(5 \cosh(4 J') -4 \cosh(2 J') - 1\right)}{16 \cosh^4(J') N} + O\left(1/N^2\right) \,,
\end{split}
\end{align}
which yields the following expression for the typical current
\begin{align}
\label{jtyp2}
j_{typ} = \frac{c_0}{2 \delta t \, \cosh^2(J') N} + O\left(1/N^2\right) \,.
\end{align}
Consequently, $j_{typ}$ appears to be smaller, due to the factor $1/\cosh^2(J') \leq 1$, than the averaged sequence-dependent current, which shows that the latter is indeed supported by anomalous realizations of disorder. 

Expression \eqref{jtyp2}  permits us to estimate, as well, the typical behavior of the sequence-dependent resistance $\tau_N$. Noticing that in virtue of the definition in eq. \eqref{j} one has that
$\ln \tau_N = - \ln(2 \delta t \, j_N/c_0))$ for any given sequence,  we find
\begin{align}
\tau_{typ} =  \exp\left(\langle \tau_N \rangle\right) = \cosh^2(J') N = \frac{\langle \tau_N\rangle}{\cosh^2(J')} \,,
\end{align}  
i.e., the typical values of $\tau_N$ are smaller than the averaged one signaling  
that
$\langle \tau_N \rangle$ is also supported by some 
 realizations of disorder which are markedly different from the typical ones.

To further highlight the difference between the typical and the averaged probability current, in Fig. \ref{fig:5}(a) we depict the typical current defined in eq. \eqref{jtyp} as function of $\beta (u_0 - u_{AB})$, confronting our theoretical prediction in eq. \eqref{jtyp2} which describes the behavior in the limit $N \to \infty$ (black solid line), and the results of numerical simulations for finite values of $N$ (see the inset).  We observe that the expression \eqref{jtyp2} provides a very accurate estimate of the typical current for any value of $\beta (u_0 - u_{AB})$ already for relatively short chains. 
Next, in Fig. \ref{fig:5}(b) to (d) we present results of the numerical averaging of the steady-state probability 
 current  in chains with a randomly-rarified binary alphabet. Here we plot $N \langle j_N\rangle$ as function of 
$\beta (u_0 - u_{AB})$  for the concentrations of voids $p = 0.05$ (b), $p = 0.1$ (c) and $p = 0.3$ (d).
   We observe that, regardless of the value of $p$, the first moment of the current tends to $J'$-independent constant when $N \to \infty$, which is exactly the behavior  in eq. \eqref{fick} implying that also for randomly-rarified binary sequences 
   the averaged current does not  "feel" the disorder.  For finite $N$, on the contrary, $N \langle j_N\rangle$ shows a noticeable variation with $\beta (u_0 - u_{AB})$, being an asymmetric function of this variable. Curiously enough, for finite $N$ the first moment of the stationary sequence-dependent current depends on $\beta (u_0 - u_{AB})$ in a completely different way as compared to the typical current. In particular, the latter is \textit{maximal} for $J' = 0$ and decreases when $\beta (u_0 - u_{AB}) \to \pm \infty$, which agrees with the common sense. On the contrary, for finite $N$ the averaged current has a deep at $J' =0$, i.e., is \textit{minimal} for $J' =0$, and tends to constant values  when $\beta (u_0 - u_{AB}) \to \pm \infty$. This signifies that the first moment of the stationary sequence-dependent current is supported by \textit{atypical} realizations of disorder also for randomly-rarified binary sequences.

\subsection{Sequence-dependent splitting probabilities.}

The splitting probability $E_-$ is  the probability that a hernia, that starts at time moment $0$ at site $x=x_0$ and moves randomly on a finite interval $(0,N)$, hits the left boundary $x = 0$ of this interval first, without ever touching the right boundary\footnote{The complementary splitting probability $E_+$, i.e., the probability that the right boundary is reached first without ever touching the left boundary, obeys $E_+ \equiv 1 - E_-$ and hence, can be trivially deduced from our results for $E_-$. } at $x = N$ (see, e.g., \cite{sidgen}).  The sequence-dependent probability $E_-$ can be expressed through the resistances $\tau_{x_0}^{(-)}$ and $\tau_{N - x_0}^{(+)}$
of the intervals $(0,x_0)$ and $(x_0,N)$ to the left and to the right from the starting point, respectively, as follows (see, e.g., \cite{sid})
\begin{align}
\label{Eminus}
E_- = \frac{\tau_{N - x_0}^{(+)}}{\tau_{N - x_0}^{(+)} + \tau_{x_0}^{(-)}} \,.
\end{align} 
Note that whenever the interval to the right from the starting point $x =  x_0$ contains an impermeable barrier or an infinitely deep trap, $\tau_{N - x_0}^{(+)} = \infty$ and hence, $E_- = 1$, as it should. 

To calculate the first two moments of the sequence-dependent probability $E_-$, we introduce an auxiliary function
\begin{align}
\Omega = \exp\left(- (\lambda_1 + \lambda_2) \tau_{N - x_0}^{(+)} - \lambda_2 \tau_{x_0}^{(-)}\right) \,,
\end{align}
and assume that $\tau_{x_0}^{(-)}$ and $\tau_{N - x_0}^{(+)}$  are uncorrelated random variables. 
In reality, they are coupled by two spin variables $\sigma_x$ in the immediate vicinity of the starting point $x = x_0$, 
but we 
discard these correlations assuming that they may incur only insignificant corrections. Consequently, we have
\begin{align}
\left \langle \Omega \right \rangle =  \Phi_{N-x_0}(\lambda_1+\lambda_2)  \Phi_{x_0}(\lambda_2) \,,
\end{align}
where the functions $\Phi$ are defined in eq. \eqref{Phi1}. The first two moments of the sequence-dependent splitting probability $E_-$ can be formally represented  using the functions $\Phi$ as
\begin{align}
\begin{split}
\left \langle E_- \right \rangle  & =  \int^{\infty}_0 d\lambda_2  \, \Phi_{x_0}(\lambda_2) \left.\left(-\frac{\partial}{\partial \lambda_1} \Phi_{N-x_0}(\lambda_1+\lambda_2)\right)\right|_{\lambda_1=0} \,, \\
\left \langle E^2_- \right \rangle &= \int^{\infty}_0 \lambda_2 \,  d\lambda_2 \, \Phi_{x_0}(\lambda_2) \left. \left( \frac{\partial^2}{\partial \lambda_1^2}  \Phi_{N-x_0}(\lambda_1+\lambda_2)\right)\right|_{\lambda_1=0} \,.
\end{split}
\end{align}
Taking advantage of the explicit result for the large-$N$ asymptotic form of $\Phi_N(\lambda)$ derived in \ref{C}, we find, after some rather tedious but straightforward calculations, the following asymptotic expression for the averaged splitting probability $E_-$: 
\begin{align}
\begin{split}
\label{Emean}
\left \langle E_- \right \rangle & = (1 - \alpha) \Big[\frac{3}{8} + \frac{\left(1- \alpha + \alpha \cosh(2 J')\right)}{2 \left((1-\alpha)^2 + \alpha^2 + 2 \alpha (1- \alpha) \cosh(2 J')\right)}\\
&+ \frac{\left(1- \alpha + \alpha \cosh(4 J')\right)}{8 \left((1-\alpha)^2 + \alpha^2 + 2 \alpha (1- \alpha) \cosh(4 J')\right)} \Big] + O\left(1/N\right) \,, \quad \alpha = x_0/N \,,\\
\end{split}
\end{align}
as well as a rather lengthy expression for the second moment of $E_-$, which we do not present here in an explicit form.  
We only discuss below its asymptotical behavior in the limits when the starting point is close to either of the boundaries and use 
the full form stored in Mathematica to test the self-averaging properties of $E_-$. Note that our theoretical results are only valid in the limit $N \gg x_0 \gg 1$, meaning that the relative distance $\alpha$ to the left boundary 
is strictly bounded away from $0$ and $1$. 

 \begin{figure}[ht]
\centering
\includegraphics[width=150mm]{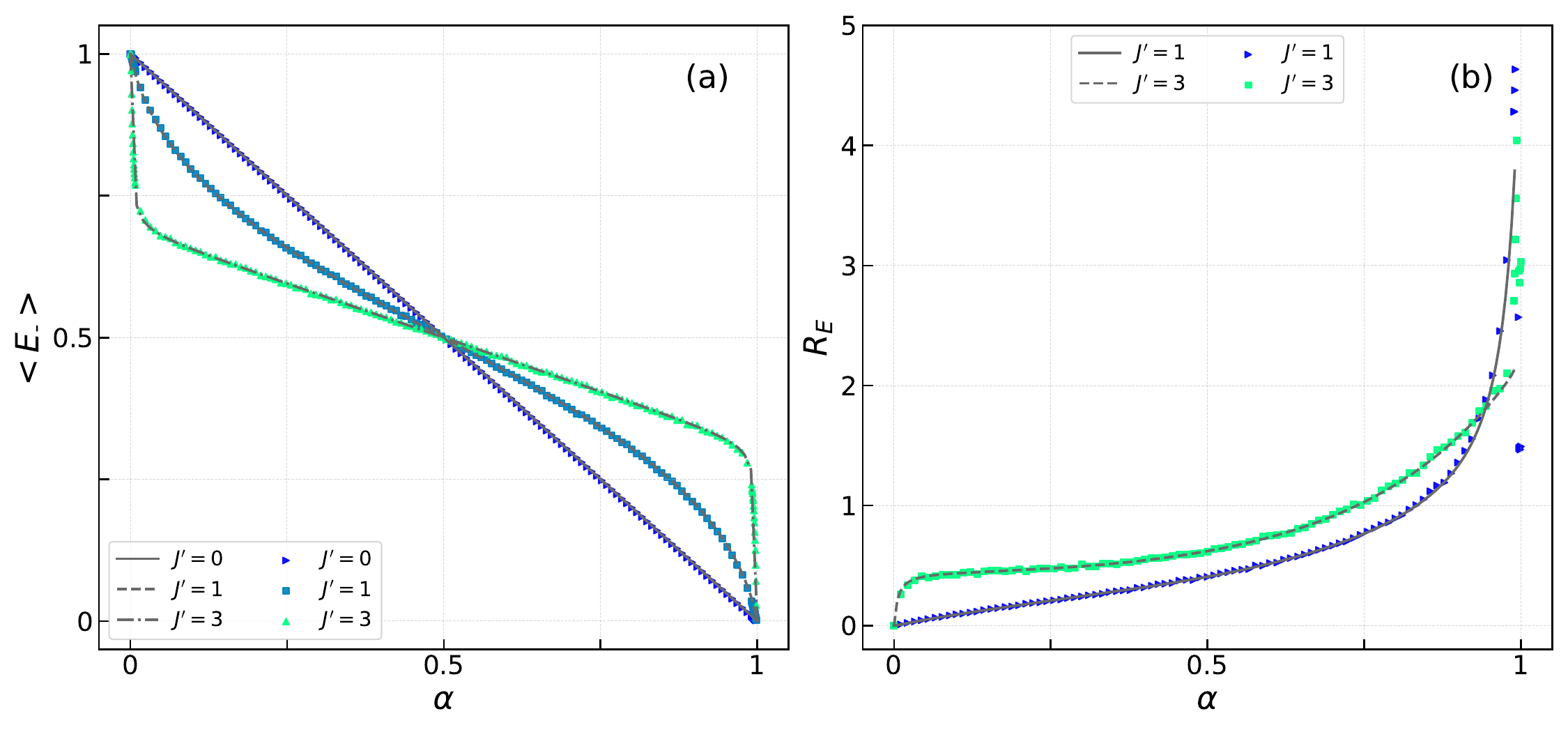} 
\caption{Splitting probability $E_-$. Panel (a): First moment of the splitting probability $E_-$ as function of the 
relative distance $\alpha = x_0/N$ to the left boundary from the starting point for $J'=0$ ($E_- \equiv 1 - \alpha$ \cite{sidgen}, solid black line), $J' = 1$ (eq. \eqref{Emean} with omitted terms $O(1/N)$, dashed black curve) and
$J' = 3$ (eq. \eqref{Emean} with omitted terms $O(1/N)$, dash-dotted  black curve). Symbols present the results of numerical averaging of expression \eqref{Eminus} over $2 \times 10^4$ realizations of sequences of letters with $N = 1024$.
Panel (b): Relative variance $R_E = \langle E^2_-\rangle/\langle E_-\rangle^2 - 1$  as function of $\alpha$. 
Solid and dashed curves define our theoretical prediction for $J' = 1$ and $J' = 3$, respectively.
Symbols present the results of a numerical averaging over $2 \times 10^4$ sequences of length $N = 1024$. Note that the numerical data shows a very irregular behavior in the vicinity of $\alpha = 1$. 
}
\label{fig:6} 
\end{figure}

We dwell first on the behavior of  $\langle E_-\rangle$ as function of the relative distance $\alpha = x_0/N$ to the left boundary and of the dimensionless interaction parameter $J'$ comparing our theoretical prediction in eq. \eqref{Emean}
(with omitted terms $O(1/N)$) and the results of a numerical averaging of the expression \eqref{Eminus}. This comparison is presented in 
Fig. \ref{fig:6}(a). We realize that our theoretical result is in a perfect agreement with the numerical data, even in the immediate vicinity of $\alpha = 0$ and $\alpha = 1$, at which points our analytical result  is not supposed to work well. 

Inspecting next our eq. \eqref{Emean}, we infer that  $\langle E_-\rangle$ is an \textit{even} function of $J'$ and that 
$\langle E_-\rangle \equiv 1 - \alpha$ for $J' = 0$ (solid black line in Fig. \ref{fig:6}(a)) - the classical result for the splitting probability $E_-$ in a homogeneous system (see \cite{sidgen}).  For arbitrary $J'$ the first moment of the splitting probability $\langle E_-\rangle \equiv 1/2 $ (and hence, $\langle E_+\rangle \equiv 1/2 $) when $\alpha = 1/2$, 
i.e., when the hernia starts right in the middle of the interval. This is, of course, not a counter-intuitive result in view of the symmetry. Further on, inspecting the expression \eqref{Emean}, we find that $\langle E_-\rangle$  approaches $1$, as it should, when $\alpha \to 0$, 
\begin{align}
\left \langle E_- \right \rangle = 1 - \cosh^4(J') \alpha + O\left(\alpha^2\right) \,,
\end{align}
and vanishes when $\alpha \to 1$,
\begin{align}
\left \langle E_- \right \rangle = \cosh^4(J') (1-\alpha) + O\left((1-\alpha)^2\right) \,.
\end{align}
Note that the coefficients in front of the leading terms becomes large when $J'$ is large signifying that the approach to the limiting values may become very abrupt, as it is indeed apparent in Fig. \ref{fig:6}(a) for $J' = 3$. 
Overall, we observe that with the growth of $J'$ the presence of disorder progressively \textit{weakens} the dependence of  $\left \langle E_- \right \rangle$ on $\alpha$; in fact, $\left \langle E_- \right \rangle$  tends to $1/2$ for any value of $\alpha$ away from $\alpha = 0$ and $\alpha = 1$. 

We turn lastly to the behavior of the relative variance $R_E = \langle E^2_-\rangle/\langle E_-\rangle^2 - 1$ which we depict in Fig. \ref{fig:6}(b) as function of the relative distance $\alpha$ for two values of the parameter $J'$. We observe  again a very good agreement between our theoretical prediction (solid and dashed curves) and the numerical data (symbols) for all values of $\alpha$ on the interval $0 \leq \alpha < 1$, including $\alpha = 0$ but excluding the immediate vicinity of $\alpha = 1$, for which a significant scatter of the numerical data is observed. 
Recall that our analytical approach is based on the assumption that $0 < \alpha < 1$, such that it is somewhat surprising that it provides a correct behavior for $\alpha = 0$ (likewise it was observed above
for the behavior of the first moment of $E_-$) and it is not, in fact, surprising that its prediction is 
in an apparent contradiction with the numerically-observed behavior in the vicinity of $\alpha = 1$.  Overall, we see that $R_E$ is greater than zero for all values of $\alpha$, except for $\alpha = 0$, which means that the splitting probability is \textit{not self-averaging}
except for the trivial case when the hernia starts right at the left boundary, for which $E_- \equiv 1$ and therefore, is not fluctuating. On symmetry grounds, it is also clear however that $E_- \equiv 0$ and is not fluctuating when the hernia starts at the right boundary, i.e., when $\alpha = 1$. Therefore, we expect a discontinuous behavior of $R_E$ at $\alpha = 1$, which apparently explains the irregular behavior of the numerical data in the vicinity of this point.     

Inspecting our analytical prediction for $R_E$, we find that in the limit $\alpha \to 0$,
\begin{align}
\label{RE}
R_E &= \frac{\sinh^2(
   J')\left(26 + 65 \cosh(2 J') + 22 \cosh(4 J') + 15 \cosh(6 J')\right) }{32} \alpha^2 + O\left(\alpha^3\right) \,,
\end{align}
which shows that $R_E$ vanishes in proportion to $\alpha^2$ when $\alpha \to 0$. Note, however, that the amplitude of the leading term becomes very large when $J'$ is large such that the rise of $R_E$ with $\alpha$ can be very steep. We indeed observe such a behavior in Fig. \ref{fig:6}(b) already for $J' = 3$. Further on, we have that when 
$\alpha$ is close to $1$, the relative variance of the splitting probability behaves as
\begin{align}
\begin{split}
\label{REE}
R_E & =  \left(\frac{\cosh^4(2 J')}{\cosh^8(J')} - 1\right) - A(J') (1-\alpha) + O\left((1-\alpha)^2\right)  \,,\\
A(J') &= \frac{\tanh^2(J')}{16 \cosh^{10}(J')}  \Bigg(22 + 27 \cosh(2 J') + 38 \cosh(4 J') + 15 \cosh(6 J')\\ &+ 16 \cosh(8 J') + 
 6 \cosh(10 J') + 4 \cosh(12 J')\Bigg) \,,
 \end{split}
\end{align}
which expression holds for any value of $\alpha$ arbitrarily close to $1$ but strictly less than it, while $R_E \equiv 0$ for $\alpha = 1$.  We verified by numerical simulations, taking larger values of $N$ and more extensive statistical samples, that the expression \eqref{REE} is indeed valid arbitrarily close to $\alpha = 1$. 

\section{Sequence-dependent mean first-passage time through a finite interval}
\label{sec:4}

Consider a finite chain with $N+2$ sites. On this chain we place an impermeable barrier at $x = -1$ (such that $p_{0,-1}=0$ and hence, $p_{0,1}=1$) and a target at the site $x=N$. 
Supposing, for simplicity, in this only Section that 
the hernia performs a hopping motion in a \textit{discrete} time, we take advantage of the well-established result for the mean first-passage time 
to the target (see, e.g., \cite{dous,nosk}). 
For the hernia that  starts at time moment $0$ from the site $x=0$, the mean  first-passage time to the target at $x = N$ is given explicitly, for an arbitrary set of transition probabilities, by \cite{dous,nosk} 
\begin{align}
\label{T}
T_N = \sum_{x=0}^{N-1} \frac{1}{p_{x,x+1}} \sum_{k=x}^{N-1} \prod_{j=x+1}^{k} \frac{p_{j,j-1}}{p_{j,j+1}} \,,
\end{align}
with the convention
\begin{align}
\prod_{j}^{k} \frac{p_{j,j-1}}{p_{j,j+1}} = 1 \,, \quad k < j \,.
\end{align}
An ample discussion of the first-passage problems in different geometries  can be found in \cite{sidgen,sidgen2,sidgen3}. 

To get an idea of the mathematical structure of expression \eqref{T}, it might be instructive to write it down explicitly
\begin{align}
\begin{split}
\label{T1}
T_N &= \frac{1}{p_{0,1}} \left[1+ \frac{p_{1,0}}{p_{1,2}} + \frac{p_{1,0} p_{2,1}}{p_{1,2} p_{2,3}} + \ldots + \frac{p_{1,0} p_{2,1} \ldots p_{N-1,N-2}}{p_{1,2} p_{2,3} \ldots p_{N-1,N}} \right] \\
&+\frac{1}{p_{1,2}} \left[1+ \frac{p_{2,1}}{p_{2,3}} + \frac{p_{2,1} p_{3,2}}{p_{2,3} p_{3,4}} + \ldots + \frac{p_{2,1} p_{3,2} \ldots p_{N-1,N-2}}{p_{2,3} p_{3,4} \ldots p_{N-1,N}} \right] \\
&+\frac{1}{p_{2,3}} \left[1+ \frac{p_{3,2}}{p_{3,4}} + \frac{p_{3,2} p_{4,3}}{p_{3,4} p_{4,5}} + \ldots + \frac{p_{3,2} p_{4,3} \ldots p_{N-1,N-2}}{p_{3,4} p_{4,5} \ldots p_{N-1,N}} \right] \\
&+ \ldots + \frac{1}{p_{N-3,N-2}} \left[1 + \frac{p_{N-2,N-3}}{p_{N-2,N-1}} + \frac{p_{N-2,N-3} p_{N-1,N-2}}{p_{N-2,N-1} p_{N-1,N}} \right]  \\
&+  \frac{1}{p_{N-2,N-1}}  \left[1 + \frac{p_{N-1,N-2}}{p_{N-1,N}} \right] + \frac{1}{p_{N-1,N}} \,,
\quad p_{0,1} = 1 \,,
\end{split}
\end{align}
which makes apparent its connection with the variable $\tau_N$ in eq. \eqref{tau} and therefore, permits to interpret $T_N$ as the sum (over all $x$) of the resistance of the interval from $x$ to the location of the target, divided by the probability that the hernia being at $x$  jumps towards this interval.

\begin{figure}[ht]
\centering
\includegraphics[width=150mm]{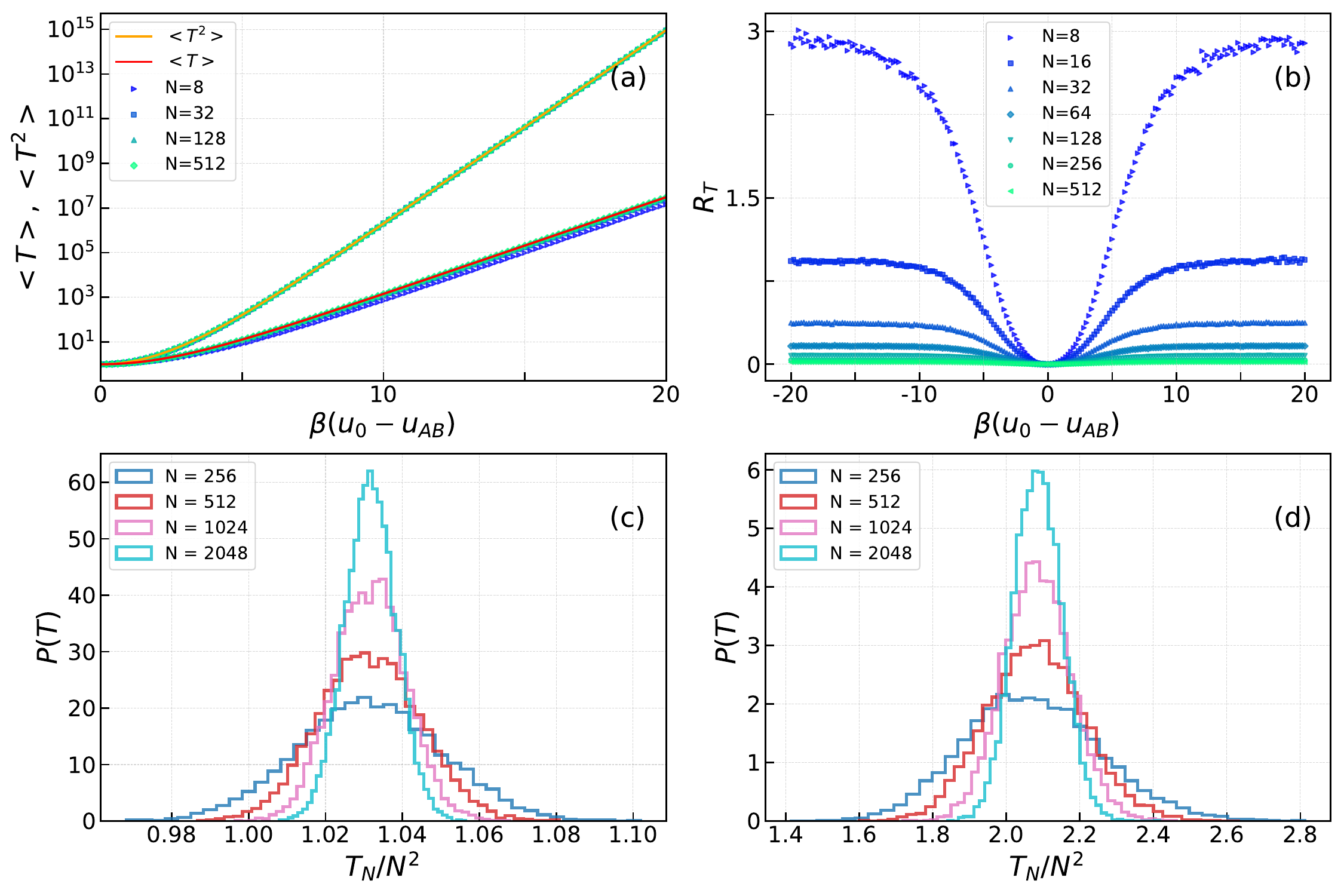} 
\caption{Mean first-passage time $T_N$ through a finite interval. Panel (a): The first and second moments of the rescaled mean first-passage time $T = T_N/N^2$; 
$\langle T\rangle$ (eq. \eqref{averagedT}, red solid curve) and $\langle T^2\rangle$ (eq. \eqref{averagedT2}, yellow solid curve) as functions of $\beta (u_0 - u_{AB})$ for four values of $N$. 
Symbols present the results of a numerical averaging of expression \eqref{T2} over $2 \times 10^4$ samples for different values of $N$ (see the inset). Panel (b): Results of numerical analysis of the relative variance $R_T = \langle T_N^2\rangle/ \langle T_N\rangle^2 - 1$ as function of $\beta (u_0 - u_{AB})$ for different values of $N$ (see the inset).
Panel (c): Numerically-evaluated probability density function $P(T)$ of the rescaled variable $T = T_N/N^2$ for 
$\beta (u_0 - u_{AB}) = 0.5$ and four different values of $N$ (see the inset). Panel (d): The same for $\beta (u_0 - u_{AB}) = 2.5$.
}
\label{fig:7} 
\end{figure}

Next, taking advantage of eqs. \eqref{33} and \eqref{phi}, we formally rewrite expression \eqref{T1} in terms of variables $\phi_x$:
\begin{align}
\begin{split}
\label{T2}
T_N & = \frac{1}{p_{0,1} \phi_0 \phi_1} \left[\phi_0 \phi_1 + \phi_1 \phi_2 + \phi_2 \phi_3 + \ldots +\phi_{N-1} \phi_N\right] \\
&+  \frac{1}{p_{1,2} \phi_1 \phi_2} \left[\phi_1 \phi_2 + \phi_2 \phi_3 + \phi_3 \phi_4  + \ldots +\phi_{N-1} \phi_N\right] \\
&+  \frac{1}{p_{2,3} \phi_2 \phi_3} \left[\phi_2 \phi_3 + \phi_3 \phi_4 + \phi_4 \phi_5  + \ldots +\phi_{N-1} \phi_N\right] \\
&+ \ldots + \frac{1}{p_{N-3,N-2} \phi_{N-3} \phi_{N-2}} \left[\phi_{N-3} \phi_{N-2} + \phi_{N-2} \phi_{N-1} + \phi_{N-1} \phi_N\right] \\
&+  \frac{1}{p_{N-2,N-1} \phi_{N-2} \phi_{N-1}} \left[\phi_{N-2} \phi_{N-1} + \phi_{N-1} \phi_N \right] +  \frac{1}{p_{N-1,N} \phi_{N-1} \phi_{N}} \phi_{N-1} \phi_N \,,
\end{split}
\end{align} 
where $p_{x,x+1}$ (except for $p_{0,1}$ which is set equal to unity) are expressed through $\phi_x$ in eq. \eqref{phi} such that the factors in front of the terms in square brackets obey 
\begin{align}
\frac{1}{p_{x,x+1} \phi_x \phi_{x+1}} = \frac{1}{\phi_{x-1} \phi_x} + \frac{1}{\phi_x \phi_{x+1}} \,.
\end{align}
Inspecting next the sum in eq. \eqref{T2} that consists of $N (N+1)/2$ terms, one finds that the disorder-averaged mean first-passage time is given, in the leading in $N$ order, by
\begin{align}
\begin{split}
\label{averagedT}
\Big \langle T_N \Big \rangle &= \left \langle \left(\frac{1}{\phi_{x-1} \phi_x} + \frac{1}{\phi_x \phi_{x+1}}\right)\right \rangle \langle \phi_x \phi_{x+1}\rangle \frac{N^2}{2} + O\left(N\right) \\
& = \cosh^4(J') N^2 + O(N)  \,,
\end{split}
\end{align}
while the second moment of $T_N$ obeys
\begin{align}
\begin{split}
\label{averagedT2}
\Big \langle T^2_N \Big \rangle &= \left \langle \left(\frac{1}{\phi_{x-1} \phi_x} + \frac{1}{\phi_x \phi_{x+1}}\right)\right \rangle^2 \langle \phi_x \phi_{x+1}\rangle^2 \frac{N^4}{4} + O\left(N^3\right) \\
& = \cosh^8(J') N^4 + O\left(N^3\right)  \,.
\end{split}
\end{align}
Note that the disorder-averaged mean first-passage time  $\langle T_N \rangle$ is proportional to $N^2$ in the leading in the large-$N$ limit order, i.e., shows exactly the same $N$-dependence as its counterpart for a random walk in a homogeneous system. A salient feature here is that in a disordered system under study $\langle T_N \rangle$ acquires a very strong dependence on the interaction energy $J'$, which signifies that the presence of disorder, regardless of the sign of $J'$,  increases the mean first-passage time making it more difficult to find the target.
Expressions \eqref{averagedT} and \eqref{averagedT2} are plotted in Fig. \ref{fig:7}(a) as functions of $\beta( u_0 - u_{AB})$ for different values of $N$. Comparison with a direct numerical averaging of the formal expression \eqref{T2} shows that the above asymptotic expressions provide very accurate estimates of the first two moments of the mean first-passage time already for quite moderate values of $N$ and also that their dependence on $J'$ is correct.

Calculation of the sub-dominant terms in eqs. \eqref{averagedT} and \eqref{averagedT2} is a very laborious task and we do not perform it here. However, considering only the 
 leading in the limit $N \to \infty$ behavior  permits us to make important conclusions. We note first that the coefficient in front of $N^4$ in eq. \eqref{averagedT2} is exactly the squared  coefficient  in front of $N^2$ in eq. \eqref{averagedT}, which signifies that the variance of $T_N$ 
 is of order $O(N^3)$ due to a cancellation of the leading terms. This agrees with the main result of \cite{slut1} which insightful work studied fluctuations of the sequence-dependent mean first-passage time for diffusion of a protein along a DNA with a disordered correlated potential. 
 In turn, such a behavior of the variance implies that the relative variance - the focal property of our present analysis - obeys
 \begin{align}
 \label{RRT}
 R_{T} = O\left(1/N\right) \,,
 \end{align}
 meaning that the mean first-passage time is \textit{strongly self-averaging} in the limit $N \to \infty$.  As we have remarked above, this result is conceptually important because it indicates the trend, showing that fluctuations become progressively less pronounced the farther the target is placed away from the starting point. 
 
 On the other hand, in most of practically important situations the target is located not far from the starting point so that one will observe a significant scatter of values of the mean first-passage times calculated on different chains. 
 We plot numerically-evaluated relative variance $R_{T}$ as function of $\beta (u_0 - u_{AB})$ in Fig. \ref{fig:7}(b). We observe that $R_T$ is a symmetric function of the coupling parameter $J'$ and may attain quite significant values for sufficiently large values of this parameter 
 and moderate values of $N$. 
 Next, in panels (c) and (d) in Fig. \ref{fig:7} we plot the numerically-evaluated full probability density function (with respect to different realizations of disorder) of the reduced mean first-passage time $T_N/N^2$ for  $\beta (u_0 - u_{AB}) = 0.5$ (c) and $\beta (u_0 - u_{AB}) = 2. 5$ (d). We observe that the probability density function is uni-modal, (in contrast to the behavior of its counterpart for the resistance $\tau_N$) and moreover, its variance diminishes when $N$ increases, which explains why $T_N$ is self-averaging when $N \to \infty$. For finite $N$ the relative variance is evidently finite and may be larger than $1$.  

\section{Sequence-dependent diffusion coefficient}
\label{sec:5}

Determining the effective 
diffusion coefficient $D$ for random motion 
in one-dimensional \textit{infinite} systems 
with a quenched random potential $U_x$ is a long-standing challenge for theorists. For continuous-space systems, one of physically plausible approaches is as follows: Consider first a standard diffusion of a particle 
on a finite interval $(0,L)$ in the presence of a given
 potential 
$U_x$, $x$ being a continuous variable, 
imposing a periodic boundary condition and also supposing that $U_x = U_{x+L}$. 
In such a setting, 
the periodicity ensures that the particle's dynamics converges to standard Brownian motion 
in the limit $t \to \infty$ (see, e.g., \cite{dean2} for a discussion of the transient behavior) 
with 
 $L$-dependent 
diffusion coefficient $D_L$ defined by 
 the 
Lifson-Jackson formula \cite{lifson} :
\begin{align}
\label{LJ}
D_L = \dfrac{D_0 L^2}{\Big(\int^L_0 dx \, e^{-\beta U_x}\Big) \Big(\int^L_0 dy \, e^{\beta U_y}\Big)} \,,
\end{align}
where  $D_0$ is the diffusion coefficient in a system without an external potential; in our settings, one has $D_0 = 1/(2 \delta t)$. Note that in eq. \eqref{LJ} the integral $\tau_L = \int^L_0 dx \, e^{-\beta U_x}$ 
is the continuous-space analogue of the resistance of the interval $(0,L)$, while the inverse of it, $j_L = D_0/\int^L_0 dx \, e^{-\beta U_x}$, is the steady-state probability current through the interval $(0,L)$. 
Respectively, the integral $\int^L_0 dy \, e^{\beta U_y}$ and its reciprocal value are the resistance and the current (up to  the multiplier  $D_0$) in a system with $U_x$ replaced by $-U_x$. Note that $D_L/D_0$ is an even function of $\beta$ and does not change upon replacement $U_x \to - U_x$.

The next step consists in averaging the expression \eqref{LJ} with respect to realizations of the potential. Finally, assuming that the limits $t \to \infty$ and $L \to \infty$ commute, one takes the limit $L \to \infty$ to find the disorder-averaged 
diffusion coefficient $\langle D \rangle$ in an infinite system. Definitely, the assumption that two limits commute is not always true, especially in situations when disorder entails \textit{anomalous} diffusion. In particular, for the so-called Sinai diffusion (see \cite{sinai,enzo2}),  $\langle D_L \rangle \to 0$ as a stretched-exponential function of $L$ in the limit $L \to \infty$ \cite{cecile}, such that one has to determine instead $\langle \ln D_L \rangle$, in order to define correctly the temporal evolution of the disorder-averaged mean-squared displacement \cite{dean}. For the case at hand, however, in view of the above-discussed behavior of the resistance, the current and the mean first-passage time, which all show exactly the same $N$-dependences as their counterparts for 
 random walks in homogeneous systems and only the amplitudes are fluctuating, 
we 
believe that such a scenario is justified and the limit $\langle D \rangle = \lim_{L \to \infty} \langle D_L\rangle$ evidently exists.

Hopping motion of the hernia studied here is defined on a discrete chain. To this end, 
we will make use of the expression for a realization-dependent diffusion coefficient $D_N$ derived four decades ago by Derrida \cite{derrida} who studied a random hopping motion in a continuous time on a  
\textit{periodic} chain with $N$ sites $x$ ($x = 1, 2, 3, \ldots, N$) and 
with \textit{arbitrary} fixed hopping rates. This expression pertains to a most general setting and has quite a complicated mathematical form. We resort therefore to the subsequent analysis made in \cite{sasha}, in which the Derrida's expression was  formally rearranged to cast it into the form which resembles the Lifson-Jackson formula \eqref{LJ}. The result for the diffusion coefficient on a finite periodic chain derived in \cite{sasha} reads 
\begin{align}
\label{sasha}
D_N = \dfrac{\alpha_1 N^2}{\left(1+ \sum_{x = 2}^{N} \prod_{j=2}^x \frac{\alpha_{j-1}}{\beta_j} \right)\left(1+ \sum_{x = 2}^{N} \prod_{j=2}^x \frac{\beta_{j}}{\alpha_j} \right)} \,,
\end{align}
where $\alpha_x$ and $\beta_x$ are the forward and backward, respectively, hopping rates from the site $x$. These rates obey $\alpha_x = \alpha_{x+N}$ and $\beta_x = \beta_{x+N}$, i.e.,  are periodic functions of $x$.

Adopting the definition of the rates 
in eq. \eqref{rates} and
taking advantage of our eqs. \eqref{33}, we find that the two factors in the denominator in eq. \eqref{sasha} follow
\begin{align}
\begin{split}
\left(1+ \sum_{x = 2}^{N} \prod_{j=2}^x \frac{\alpha_{j-1}}{\beta_j} \right) &= 2 \, \alpha_1 \, \phi_1 \, \phi_2 \, \delta t \sum_{x=1}^N \frac{1}{\phi_x \phi_{x+1}} \,, \\
\left(1+ \sum_{x = 2}^{N} \prod_{j=2}^x \frac{\beta_{j}}{\alpha_j} \right) &= \frac{1}{\phi_1 \phi_2} \sum_{x=1}^{N} \phi_x \phi_{x+1} \,,
\end{split}
\end{align}
with $\phi_{N+1} = \phi_1$. Consequently, for our model 
the expression in eq. \eqref{sasha} attains the following exact and compact form
\begin{align}
\begin{split}
\label{DD}
D_N &= \frac{N^2}{2 \delta t \left(\sum_{x=1}^N \dfrac{1}{\phi_x \phi_{x+1}}\right)\Big(\sum_{x=1}^{N} \phi_x \phi_{x+1} \Big)} \\
&=\frac{N^2}{2 \delta t \, \tau_N(J) \tau_N(-J)} \,,
\end{split}
\end{align}
where $\phi_{N+1} = \phi_1$,  $\sigma_{N+1} = \sigma_1$, $\tau_N(J)$ is the resistance of a 
chain with $N$ sites $x = 1, 2, 3, \ldots, N$ (compare with eq. \eqref{taup}), while $\tau_N(-J)$ is the resistance of the same chain with the coupling constant $J$ replaced by $-J$.   Note that in the absence of disorder one finds from eq. \eqref{DD} that $D_N = 1/(2 \delta t)$, i.e., the standard expression for the diffusion coefficient of a continuous-time hopping motion on a lattice of integers. 

\begin{figure}[ht]
\centering
\includegraphics[width=150mm]{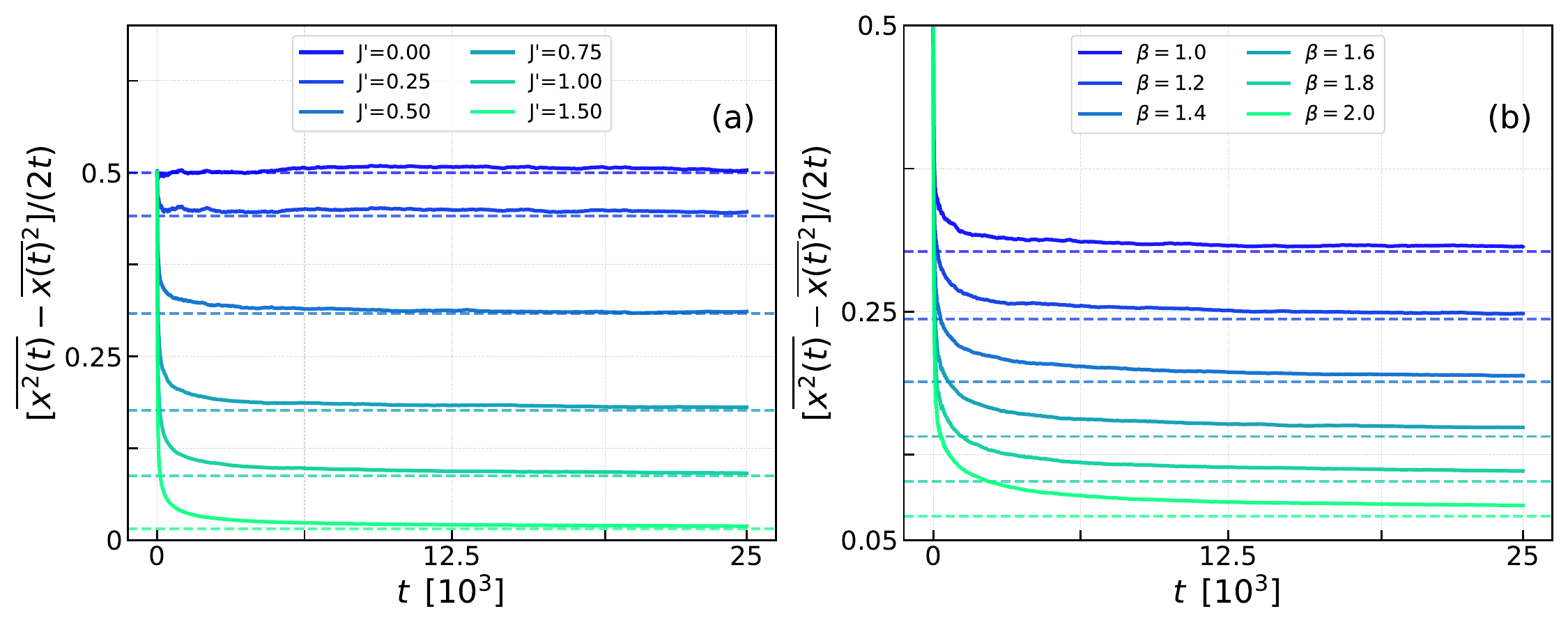}  
\caption{Sequence-dependent diffusion coefficient in a periodic system. Noisy solid curves depict the time-evolution of the ratio $(\overline{X^2(t)} - \overline{X(t)}^2)/(2 t) (= D_N(t))$ evaluated in numerical simulations of a continuous-time hopping motion on a periodic chain with $N=1024$ sites in presence of a frozen disorder.   
Thermal averaging is performed over $5 \times 10^4$ trajectories $X(t)$ starting from random positions.
Panel (a): Check of the convergence to the expression \eqref{DD} for a fixed randomly-generated sequence of letters $A$ and $B$ for several values of $J'$ (see the inset).  
Dashed horizontal lines define the realization-dependent diffusion coefficient in eq. \eqref{DD} for corresponding values of $J'$.  Panel (b): Check of the convergence to the expression \eqref{DDD} in which $U_x$  are independent, identically distributed random Gaussian variables for several values of $\beta$ (see the inset). 
 Dashed horizontal lines define the realization-dependent diffusion coefficient in eq. \eqref{DDD} for 
 such a choice of $U_x$ and corresponding values of $\beta$.
 }
\label{fig:8} 
\end{figure}

We have performed numerical simulations of the hopping process with the rates in eq. \eqref{rates}  on a periodic lattice of integers for a fixed randomly-generated sequence of letters. 
We recorded individual trajectories $X(t)$ of the hernia and determined the time evolution of the
variance $\overline{X^2(t)} - \overline{X(t)}^2$, thermal averaging being performed over $5 \times 10^4$ trajectories starting at randomly chosen points.  
The time-evolution of the ratio $(\overline{X^2(t)} -  \overline{X(t)}^2)/2 t$, (which should converge to the realization-dependent diffusion coefficient), is depicted in Fig. \ref{fig:8}(a) for several values of $J'$, and compared against  the analytical prediction in eq. \eqref{DD}.  Convergence to the result in eq. \eqref{DD} is apparent.

Before we proceed to calculation of  the moments of $D_N$ in eq. \eqref{DD}, the following important remark is in order: Expression \eqref{DD} is only valid for the explicit choice of the form of the potential $U_x$ in eq. \eqref{1}. 
Consider a more general situation in which a continuous-time hopping process evolves on a discrete \textit{periodic} 
chain with $N$ sites and at each site there is 
a on-site potential $U_x$, ($x=1, 2, \ldots, N$, $U_{x}=U_{x+N}$), which is \textit{absolutely} arbitrary.  
We stipulate next that the transition rates of such a process are given by eq. \eqref{rates} with the normalized probabilities $p(x,x+1)$ and $p(x,x-1)$, which depend on the on-site potential as it 
is defined in the first and the third lines in eqs. \eqref{3}.  Upon some straightforward algebra, we find from eq. \eqref{sasha} that for such a process the realization-dependent diffusion coefficient $D_N$ reads
\begin{align}
\label{DDD}
D_N = \dfrac{N^2}{2 \delta t \left(\sum_{x=1}^N e^{-\frac{\beta}{2}\left(U_x + U_{x+1}\right)}\right)\left(\sum_{y=1}^N e^{\frac{\beta}{2}\left(U_y + U_{y+1}\right)}\right)} \,, U_{x} = U_{x+N} \,.
\end{align}
This \textit{exact} and physically meaningful  expression matches perfectly well the Lifson-Jackson formula, 
showing, however,  that 
a "correct" discretization of the continuous-space result in eq. \eqref{LJ} is rather non-evident. Note that choosing $U_x$ as in eq. \eqref{1}, we retrieve the expression \eqref{DD}. We checked the convergence to eq. \eqref{DDD} by simulating numerically a random hopping motion process 
$X(t)$ with the transition probabilities defined in the first and the third lines in eqs. \eqref{3} on a periodic chain of integers with a given realization of the sequence of the
on-site potentials $U_x$, assuming that the latter are independent (for different $x$), identically-distributed random variables drown from the Gaussian distribution $P(U_x) = \exp(-U_x^2/2)/\sqrt{2 \pi}$. 
In Fig. \ref{fig:8}(b) we plot the ratio $(\overline{X^2(t)} - \overline{X(t)}^2)/(2 t)$ (the thermal averaging being performed over  $5 \times 10^4$ trajectories $X(t)$ starting from random positions) as function of $t$ for several values of the reciprocal temperature $\beta$. We again observe a very good convergence to the expression \eqref{DDD} noticing, however, that the larger the value of $\beta$ is, the more time is needed to thermalize the process and hence, to reach the asymptotic value of the diffusion coefficient.

We turn finally to the moments $\langle D_N^n\rangle$ of the realization-dependent diffusion coefficient in eq. \eqref{DD}.
In \ref{D} we calculate the large-$N$ asymptotic form of the  following "partition function"
\begin{align}
\label{Theta}
\Theta_N = \left \langle \exp\left( -  \frac{\lambda_1}{N} \sum_{x=1}^N \frac{1}{\phi_x \phi_{x+1}} - \frac{\lambda_2}{N} \sum_{x=1}^{N} \phi_x \phi_{x+1} \right) \right \rangle \,, \quad \lambda_1 \geq 0 \,, \lambda_2 \geq 0 \,,
\end{align}
from which the large-$N$ asymptotic of all the moments of $D_N$, not necessarily of an integer order $n > 0$,  are obtained by 
multiplying the above expression by $(\lambda_1 \lambda_2)^{n-1}$ and performing the integration
over $\lambda_1$ and $\lambda_2$; that being,  
\begin{align}
\label{momk}
\Big \langle D_N^n \Big \rangle = \frac{1}{\left(2 \delta t \right)^n \Gamma^2(n)} \int^{\infty}_0 \int^{\infty}_0 d\lambda_1 \, d\lambda_2 \, (\lambda_1 \lambda_2)^{n-1} \Theta_N \,.
\end{align} 
In doing so, we find the following general expression which is valid in the limit $N \to \infty$, 
\begin{align}
\begin{split}
\label{Dmom}
\Big \langle D_N^n \Big \rangle &= \frac{1}{\left(2 \delta t \cosh^4(J')\right)^n} \left(1 + \frac{B_n(J')}{ N} + O\left(1/N^2\right)\right) \,, \\
&B_n(J') = n \left(5+ 2n - \frac{\left(2+ (3 +2 n) \cosh(2 J')\right)}{\cosh^4(J')} \right) \,.
\end{split}
\end{align}
Note that $B_n$ (as well as all the coefficients in front of higher inverse powers of $N$, which we do not present in explicit form) vanishes when $n=0$ or when $J'=0$.

\begin{figure}[ht]
\centering
\includegraphics[width=150mm]{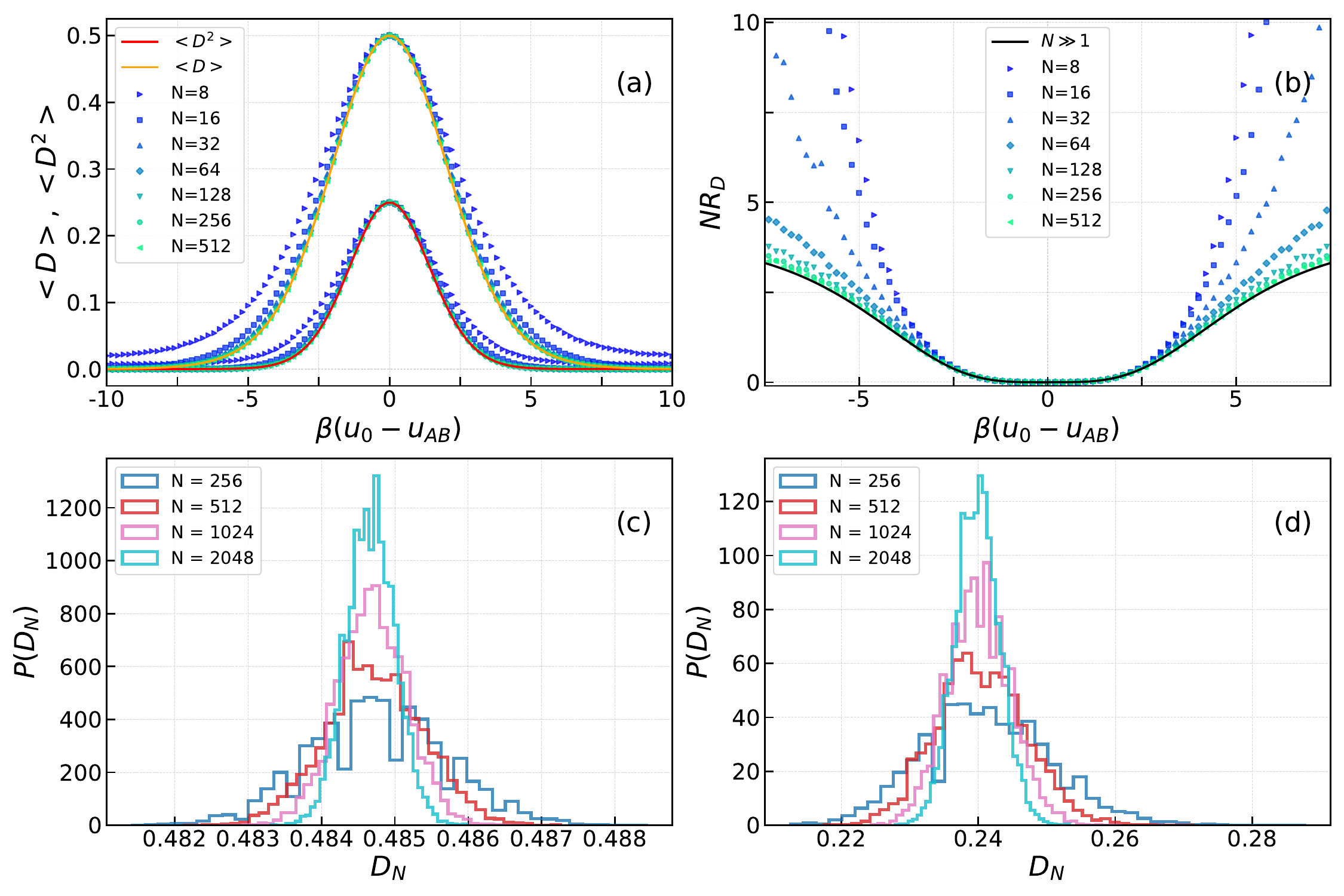}  
\caption{Sequence-dependent diffusion coefficient in a periodic system. Panel (a): 
The first and the second moments of the sequence-dependent diffusion coefficient  $D_N$  (eq. \eqref{DD}). Yellow and red solid curves define our theoretical predictions for 
$\langle D \rangle = \lim_{N \to \infty} \langle D_N \rangle$ and $\langle D^2 \rangle = \lim_{N \to \infty} \langle D_N^2\rangle$ (eq. \eqref{Dmom} with $N = \infty$).
Symbols indicate the results of a direct numerical averaging of the expression \eqref{DD} over $20000$ samples for several finite values of $N$ (see the inset). Panel (b): Relative variance $R_D$ multiplied by $N$ as function of $\beta (u_0 - u_{AB})$. Black solid curve depicts our analytical result in eq. \eqref{RD}. Symbols present numerically-evaluated relative variance for different values of $N$ (see the inset). Panels (c) and (d): Numerically-evaluated probability density function $P(D_N)$  as function of $D_N$ for $\beta (u_0 - u_{AB}) = 0.5$ and $\beta (u_0 - u_{AB}) = 2.5$, respectively, for four values of $N$.
 }
\label{fig:9} 
\end{figure}

In Fig. \ref{fig:9}(a) we depict the first two moments of the sequence-dependent random variable $D_N$ as function of $\beta (u_0 - u_{AB})$ and compare our theoretical predictions for $\langle D^n\rangle = \lim_{N \to \infty} \langle D_N^n\rangle$ with $n = 1$ and $n=2$  against the results obtained by a numerical averaging of the expression \eqref{DD} for several fixed values of $N$, observing quite a fast convergence to the limiting values upon an increase of $N$.  
The first two moments of $D_N$ defined by eq. \eqref{Dmom} entail the following result for the relative variance
\begin{align}
\label{RD}
R_D = \frac{32 \sinh^4(J')}{\left(3 + 4 \cosh(2 J') + \cosh(4 J')\right) N} + O(1/N^2) \,,
\end{align}
implying that $D_N$ is \textit{strongly} self-averaging (see the discussion below eq. \eqref{R}) in the limit $N \to \infty$.
As evidenced in Fig. \ref{fig:9}(b), 
$R_D$ remains finite, of course, for any large but finite $N$. 
Further on, in Fig. \ref{fig:9}(c) and (d) we plot a numerically-evaluated probability density function $P(D_N)$ of $D_N$ for $\beta (u_0 - u_{AB}) = 0.5$ and $\beta (u_0 - u_{AB}) = 2.5$, respectively, and for four values of $N$. We observe that, likewise the probability density function of the mean first-passage time, $P(D_N)$ is a uni-modal function and its variance vanishes as $N$ increases. 
 Such a self-averaging of $D_n$ stems apparently from the fact that in an infinite chain and during an infinite time the hernia probes all possible combinations of sequences present in an ensemble of all possible finite samples. We finally remark that our eq. \eqref{Dmom} ensures that, in the limit $N \to \infty$, the typical value of the diffusion coefficient coincides with the averaged one, i.e., 
 \begin{equation}
 D_{typ} = \exp\left(\Big \langle \ln D \Big \rangle\right) = \Big \langle D \Big \rangle \,,
 \end{equation}
 which means that in this limit the moments of the diffusion coefficient in the model under study are supported by typical realizations of disorder, in contrast to the moments of the resistance and of the steady-state sequence-dependent current.
 More arguments about self-averaging of the diffusion coefficient in an infinite system, when it exists, can be found in \cite{derrida} and \cite{aslan}.

\section{Conclusions}
\label{sec:6}

In this paper we made a first step towards a conceptual understanding of the challenging
 question about the sensitivity of a diffusive transport of knots along a given DNA to the actual sequence of nucleotides placed on it. We proposed a simple model which captures only several basic features of a real complex physical system, while discarding other factors which may potentially turn out to be quite significant. At such an expense, our model is exactly solvable permitting therefore to gain some important insight into the following two important aspects, such as   
 a)  the dependence 
 of several characteristic transport properties on the interaction energies of nucleotides and b) whether these transport properties are self-averaging with respect to different realizations of the sequences of nucleotides, or not. The latter issue is of a special interest; indeed, in a self-averaging system it is enough to have a single, sufficiently large sample and the behavior observed on this sample will be nonetheless representative of a statistical ensemble of all possible samples. In the absence of self-averaging,  on the contrary, the value of an observable deduced from 
 measurements performed on a single sample, arbitrarily large, will not give 
 a representative result and must be repeated on an ensemble of samples. Therefore, for the properties that lack self-averaging one expects to observe a significant scatter of values when the measurements are performed on different polymers.

More specifically, we considered a stretched chain carrying a frozen random sequence of letters $A$ and $B$, which  are present in the same, on average, amounts and have different interaction energies. A chain contains a single distortion - not a knot but rather a hernia (see Fig. \ref{fig:1}) - which shortens the distance between the two letters at its bottom such that they become interacting. Other letters along the chain are placed sufficiently far from each other and therefore are non-interacting.   
 Due to interactions with the solvent in which the chain is immersed, the hernia performs an activated hopping motion along the chain. Displacing itself along the chain, the hernia therefore "reads" the sequence of letters and its transition probabilities depend on the types of four letters appearing in the vicinity of its actual position.  We remark that such a model is constructed in order to mimic, in some effective way, a random translocation of a knot along the chain bearing a random sequence of nucleotides, due to either 
 a self-reptation \cite{kar,gros} or a knot region breathing \cite{metz} - essentially one-dimensional stochastic processes in which different "letters" on 
 the chain within the knot are brought in contact affecting therefore its local transition probabilities. 
 
For the model under study we concentrated on five sequence-dependent random variables characterizing 
a diffusive dynamics of the hernia: Resistance of a finite chain of length $N$ with a frozen sequence of letters with respect to a random transport;  The current through a finite chain of length $N$;  Splitting probability, i.e.,  the probability that for a diffusion on a finite chain starting from some fixed position within the chain the left end of the latter will be reached first without ever hitting its right end;  Mean first-passage time through a finite chain with $N$ sites, and finally, 
the diffusion coefficient in a periodic chain with $N$ sites.  We have determined exactly the dependences of the moments of these random, sequence-dependent variables on the interaction energies of letters. Further on, we have shown
 that the first three random variables, which are mathematically linked to each other\footnote{Note that the current is just the inverse of the resistance, while the splitting probability is a simple functional of the resistances of the intervals from the left and from the right from the hernia's starting point.}, are \textit{not self-averaging} in the limit $N \to \infty$, which signifies that these three properties will exhibit strong sample-to-sample fluctuations for any value of $N$. Additionally, we have shown that the moments of the resistance and of the current (and apparently of the splitting probability) are supported by anomalous realizations of the disordered sequences of letters and their behavior is therefore markedly different from the typical one, which we have also specified. 
 
 In turn, we have demonstrated that 
   the sequence-dependent mean first-passage time and the diffusion coefficient are both strongly self-averaging in the limit $N \to \infty$. In consequence, these properties are fluctuating for the intermediate values of $N$ but the fluctuations will become insignificant in the limit $N \to \infty$. In this limit, the disorder-averaged diffusion coefficient coincides with the typical value of the diffusion coefficient which should be observed for most of realizations.    
   We emphasize that the latter result on the asymptotic self-averaging is very meaningful for the diffusion coefficient, since it concerns the behavior which should be observed in an infinite system. Conversely, for the mean first-passage time it rather indicates the trend that fluctuations become progressively less important the farther appears the target from the starting point and vanish eventually when the target is placed at an infinite distance from the latter. However, in most of applications the target is located only at some finite distance away from the starting point and hence, fluctuations are generally non-negligible. As a consequence, for finite $N$ the disorder-averaged mean first-passage time alone will not describe the behavior 
adequately well and the knowledge of the full probability density function is desirable. 

Via numerical simulations we have studied the spread of fluctuations for finite $N$ 
presenting an evidence that 
the sample-to-sample fluctuations can be quite significant  for all the properties under study. 
Finally, we have evaluated numerically the full probability density functions of three of the five random variables under study. We have realized that the probability density function of the
 resistance of a finite chain is three-modal, which explains why this property lacks self-averaging, as well as the steady-state realization-dependent current and the splitting probability. In turn, the 
 probability density functions of the mean first-passage time and of the diffusion coefficient are both unimodal, and moreover, their variances decrease upon an increase of $N$,  which entails the self-averaging behavior in the limit $N \to \infty$. 
 
Our model can be evidently extended in several directions by considering, e.g., alphabets with more than two letters which is a more realistic situation, or by studying sequences with strong correlations in spatial placement of letters \cite{madden}. Concurrently, 
 we believe that the very fact that several interesting effects emerge 
even in the simple case studied here
will
prompt further analysis of this fascinating stochastic process of knot's diffusion along a disordered chain, stimulating 
the appearance of more sophisticated theoretical models and of Brownian dynamics simulations of knot's diffusion on DNAs which take into account explicitly the presence 
nucleotides. 
 
 \ack

We wish to thank O. B\'enichou, J.-H. Jeon,  L. Mirny, S. Nechaev, E. Orlandini and L. Pastur for helpful discussions. This project has been supported by funding from the 2021 first FIS
(Fondo Italiano per la Scienza) funding scheme (FIS783 - SMaC -
Statistical Mechanics and Complexity University and Research) from Italian MUR
(Ministry of University and Research).

\appendix

\section{Sequence-dependent probability current}
\label{A}

Consider a finite chain in which we fix the value of the probability $P_{x=0} = P(x=0,t) = c_0$ (see eq. \eqref{ME}) 
at the site $x=0$, and place a perfect trap at the site $x = N$, such that  $P_{x=N} = P(x=N,t) = 0$. 
Then, in such a situation the evolution equations \eqref{ME} become 
\begin{align}
\begin{split}
&\delta t \, \dot{P}_1=  - \left(p_{1,0} + p_{1,2}\right) P_1 + p_{0,1} c_0  + p_{2,1} P_2 \,, \\
&\delta t \, \dot{P}_2 = - \left(p_{2,1} + p_{2,3}\right) P_2 + p_{1,2} P_1  + p_{3,2} P_3 \,, \\
&\delta t \, \dot{P}_3 = - \left(p_{3,2} + p_{3,4}\right) P_3 + p_{2,3} P_2  + p_{4,3} P_4 \,, \\
&\ldots \\
&\delta t \, \dot{P}_{N-1} = -  \left(p_{N-1,N-2} + p_{N-1,N}\right) P_{N-1} + p_{N-2,N-1} P_{N-2} \,.
\end{split}
 \end{align}
Setting next $\dot{P}_x = 0$ for all $x$,  and solving the resulting linear system of equations recursively, we get for $P_{N-1}$:
\begin{align}
P_{N-1} = \frac{p_{0,1} c_0}{p_{N-1,N} \, \tau_N} \,,
\end{align}
where $\tau_N$ is defined in the main text in eq. \eqref{tau}.
Respectively, the sequence-dependent current to the trap at $x = N$ obeys
\begin{align}
j_N = \frac{p_{N-1,N} P_{N-1}}{\delta t} = \frac{p_{0,1} c_0}{\delta t \, \tau_N} \,.
\end{align}
The last step consists in doing a simplifying assumption about the value of the energy that a hernia has being at the site $x = -1$. We stipulate that $U_{-1} = U_1$ (i.e., the energy at the site $x = 1$) and hence, $p_{0,1} = 1/2$, which yields the expression \eqref{j}.

\section{Moment-generating function of the sequence-dependent resistance}
\label{C}

We seek the moment-generating function of $\tau_N$ in eq. \eqref{Phi1}, which we formally expand into the Taylor series as
\begin{equation}
\label{Cdd}
\Phi_N = \left \langle e^{- \lambda \tau_N} \right \rangle  = e^{- \lambda} \sum_{n=0} \frac{(- \lambda)^n}{n!} \Big \langle \left(\tau_N - 1\right)^n \Big \rangle   \,,
\end{equation}
We  concentrate next on the $n$-th moment of variable $\tau_N - 1$:
\begin{align}
\begin{split}
\label{Ca}
&\Big \langle \left(\tau_N - 1\right)^n \Big \rangle = \Bigg \langle \frac{\left(\phi_1 \phi_2 +  \phi_2 \phi_3 + \phi_3 \phi_4 + \ldots + \phi_{N-2} \phi_{N-1} + \phi_{N - 1} \phi_N   \right)^n }{\phi_0^n \phi_1^n} \Bigg \rangle \\
&=  \sum \frac{n!}{a_1! a_2! a_3! \ldots a_{N-1}!} \left \langle \phi_0^{b_0} \phi_1^{b_1} \phi_2^{b_2} \phi_3^{b_3} \ldots \phi_{N-1}^{b_{N-1}} \phi^{b_N}_{N}  \right \rangle  \,,
\end{split}
\end{align}
where the sum extends over all positive integer solutions of equation
\begin{align}
\label{Cm1}
a_1 + a_2 + a_3 + \ldots + a_{N- 1}  = n \,,
\end{align}
while
\begin{align}
\begin{split}
\label{Cm2}
&b_0 = - n, \\ &b_1 = - n + a_1 \,, \,\,\, b_2 = a_1 + a_2\,, \,\,\, b_3 = a_2 + a_3 \,, \ldots \,, b_{N - 1} = a_{N-2} + a_{N-1} \,, \\
 &b_N = a_{N-1} \,.
 \end{split}
\end{align}
The averaging of the expression in the second line in eq. \eqref{Ca} can be performed very directly. We start with the variable $\sigma_{N+1}$, then we average over $\sigma_{N}$, and so on. In doing so, we have 
\begin{align}
\begin{split}
\label{Cc}
&\left \langle \phi_0^{b_0} \phi_1^{b_1} \phi_2^{b_2} \phi_3^{b_3} \ldots \phi_{N-1}^{b_{N-1}} \phi^{b_N}_{N}  \right \rangle = \prod_{k=1}^N \cosh\left(J' b_k\right) \\
& = \cosh\left(n J' \right) \, \cosh\left((a_1 - n) J'\right) \, \cosh\left((a_1 + a_2) J'\right) \, \cosh\left((a_2 + a_3) J'\right) \ldots \\
& \ldots \cosh\left((a_{N-2} + a_{N-1}) J'\right) \, \cosh\left(a_{N-1} J'\right) \,.
\end{split}
\end{align}
Now, we have to perform the summation over all $a_k$ which obey the Diophantine eq. \eqref{Cm1}. For such a purpose, the expression in the second line in eq. \eqref{Cc} appears to be rather inconvenient because each $a_k$ enters simultaneously into the arguments of two hyperbolic cosines. We therefore have to decouple them which can be done by 
 introducing a set of auxiliary spin variables $s_k$, each assuming (with equal probability $1/2$) the values $\pm 1$. Taking next advantage of the equality
\begin{equation}
\cosh\left((a_{k-1} + a_{k}) J'\right) = \Big \langle \exp\Big( (a_{k-1} + a_{k}) J' s_k\Big) \Big \rangle_{s_k} \,, 
\end{equation}
we have
\begin{align}
\begin{split}
&\left \langle \phi_0^{b_0} \phi_1^{b_1} \phi_2^{b_2} \phi_3^{b_3} \ldots \phi_{N-1}^{b_{N-1}} \phi^{b_N}_{N}  \right \rangle   = \cosh\left(n J' \right) \\ &\times \left \langle e^{-n J' s_1 + (s_1 + s_2) J' a_1 + (s_2 + s_3) J' a_2 + \ldots +  (s_{N-1} + s_N) J' a_N} \right \rangle_{\{s\}} \,,
\end{split}
\end{align}
where the subscript $\{s_k\}$ signifies that averaging is performed over all auxiliary spin variables. Then, in virtue of eqs. \eqref{Cc} and \eqref{Ca}, we have that the $n$-th moment of $(\tau_N-1)$ obeys
\begin{equation}
\label{Cd}
\Big \langle \left(\tau_N - 1\right)^n \Big \rangle = \cosh\left(n J'\right) \Big \langle e^{- n J' s_1} \Big[e^{J'(s_1 + s_2)} + e^{J' (s_2 + s_3)} + \ldots + e^{J' (s_{N-1} + s_N)} \Big]^n \Big \rangle_{\{s\}}
\end{equation} 
Lastly, we use the equality
\begin{align}
e^{J' (s_{k-1} + s_k)} = \cosh^2\left(J'\right) \Big(1 + t (s_{k-1} + s_k) + t^2 s_{k-1} s_k \Big) \,, \quad t = \tanh(J') \,, 
\end{align}
which permits us to formally rewrite eq. \eqref{Cd} as
\begin{equation}
\label{Ct}
\Big \langle \left(\tau_N - 1\right)^n \Big \rangle = \cosh\left(n J'\right) \cosh^{2 n}\left(J'\right) \Big \langle e^{- n J' s_1} \Big[N - 1 - t (s_1 + s_N) + {\cal H}_N \Big]^n \Big \rangle_{\{s\}} \,,
\end{equation}
where ${\cal H}_N$ is the "Hamiltonian" of a one-dimensional Ising model with the coupling $t^2$ and
magnetic field $2 t$ :
\begin{equation}
{\cal H}_N = t^2 \sum_{k=1}^{N-1} s_k s_{k+1} + 2 t \sum_{k=1}^N s_k \,.
\end{equation}
In what follows, we focus on the limit $N \gg 1$, so that $N \gg 1 + t (s_1 + s_N)$, and discard the corresponding (insignificant) term in eq. \eqref{Ct}. 

Inserting eq. \eqref{Ct} into eq. \eqref{Cdd} and performing summation, we find that
 the moment-generating function in eq. \eqref{Cdd} is given by
\begin{equation}
\label{phiN}
\Phi_N = \frac{1}{2} e^{- \lambda} \left \langle e^{- \lambda \Omega_+ N} \left \langle e^{- \lambda \Omega_+ {\cal H}_N} \right \rangle_{\{s_k'\}}
\right \rangle_{s_1} +  \frac{1}{2} e^{- \lambda} \left \langle e^{- \lambda \Omega_- N} \left \langle e^{- \lambda \Omega_- {\cal H}_N} \right \rangle_{\{s_k'\}}
\right \rangle_{s_1}  \,,
\end{equation}
where the inner angle brackets with the subscript $\{s_k'\}$ denote averaging over all spin variables $s_k$, except for $s_1$ which is kept fixed,  the outer angle brackets with the subscript $s_1$ stand for averaging over this variable only, and $\Omega_{\pm}$ are shorthand notations for
\begin{equation}
\Omega_{\pm} = \cosh^2(J') e^{\pm J' (1 \mp s_1)} \,.
\end{equation}

We focus on the "partition" function of the form
\begin{equation}
\label{m}
{\cal Z}_N = \left \langle e^{- z {\cal H}_N} \right \rangle_{\{s_k'\}} \,.
\end{equation}
Evidently, we have
\begin{equation}
{\cal Z}_N = \frac{1}{2} {\cal Z}^{+}_N + \frac{1}{2} {\cal Z}^{-}_N \,,
\end{equation}
where the first term in the right-hand-side of the latter equation is ${\cal Z}_N$ with the spin $s_N=1$, while $ {\cal Z}^{-}_N$ is the partition function with the last spin $s_N = - 1$. It is straightforward to find that ${\cal Z}^{+}_N$ and ${\cal Z}^{-}_N$ obey the recursion 
\begin{align}
\begin{split}
\label{2}
{\cal Z}^{+}_N &= \frac{1}{2} e^{- z (2 t+t^2)} {\cal Z}^{+}_{N-1}  + \frac{1}{2} e^{-  z (2 t - t^2)} {\cal Z}^{-}_{N-1}  \,, \\
{\cal Z}^{-}_N &= \frac{1}{2} e^{z (2 t+t^2)} {\cal Z}^{+}_{N-1}  + \frac{1}{2} e^{z (2 t - t^2)} {\cal Z}^{-}_{N-1}  \,,
\end{split}
\end{align}
which is to be solved subject to the "initial" conditions
\begin{equation}
{\cal Z}^{+}_2(z,s_1) =  e^{- z ((2 t+t^2) s_1 + 2t)} \,, \quad {\cal Z}^{-}_2(z,s_1) =  e^{- z ((2 t- t^2) s_1 - 2t)} \,.
\end{equation}
One introduces next a pair of generating functions
\begin{equation}
{\cal Z}^+ = \sum_{N=2}^{\infty}  {\cal Z}^+_N \zeta^N \,, \quad  {\cal Z}^- = \sum_{N=2}^{\infty}  {\cal Z}^{-}_N \zeta^N \,,
\end{equation}
such that 
\begin{equation}
{\cal Z} = \frac{1}{2} {\cal Z}^+ + \frac{1}{2} {\cal Z}^-
\end{equation}
Multiplying both sides of eqs. \eqref{2} by $\zeta^N$ and performing summations, we have
\begin{align}
\begin{split}
\label{22}
{\cal Z}^+ &=  \zeta^2 {\cal Z}_2^+(z,s_1) + \frac{1}{2} e^{- z (2 t+t^2)} \zeta  {\cal Z}^+  + \frac{1}{2} e^{- z (2 t - t^2)} \zeta  {\cal Z}^- \,, \\
{\cal Z}^- &=  \zeta^2 {\cal Z}_2^-(z,s_1) + \frac{1}{2} e^{z (2 t+t^2)} \zeta  {\cal Z}^+  + \frac{1}{2} e^{z (2 t - t^2)} \zeta  {\cal Z}^- \,.
\end{split}
\end{align}
Solving eqs. \eqref{22} we obtain
\begin{align}
\label{ju}
{\cal Z} &=  \zeta^2 \frac{{\cal Z}^+_2 + {\cal Z}^-_2 + e^{2 z t} \sinh(z t^2) \zeta {\cal Z}^+_2 + e^{- 2 z t} \sinh(z t^2) \zeta {\cal Z}^-_2 } {2 \left(1 - e^{-z t^2} \cosh(2 z t) \zeta - \frac{1}{2} \sinh(2 z t^2) \zeta^2\right)} \nonumber\\
&= \frac{\zeta^2}{2} \left({\cal Z}^+_2(z,s_1) + {\cal Z}^-_2(z,s_1) + e^{2 z t} \sinh(z t^2) \zeta {\cal Z}^+_2(z,s_1) + e^{- 2 z t} \sinh(z t^2) \zeta {\cal Z}^-_2(z,s_1) \right) \nonumber\\
&\times  \left(\frac{\zeta_2(z)}{\zeta_2(z) - \zeta_1(z)} \frac{1}{\left(1 - \zeta/\zeta_1(z)\right)}  - \frac{\zeta_1(z)}{\zeta_2(z) - \zeta_1(z)} \frac{1}{\left(1 - \zeta/\zeta_2(z)\right)}  \right) \nonumber\\
&= \frac{\zeta^2}{2} \left({\cal Z}^+_2(z,s_1) + {\cal Z}^-_2(z,s_1) + e^{2 z t} \sinh(z t^2) \zeta {\cal Z}^+_2(z,s_1) + e^{- 2 z t} \sinh(z t^2) \zeta {\cal Z}^-_2(z,s_1) \right) \nonumber \\
&\times  \Bigg(\frac{\zeta_2(z)}{\zeta_2(z) - \zeta_1(z)} \sum_{N=0}^{\infty}\frac{\zeta^N}{\zeta_1^N(z)}  - \frac{\zeta_1(z)}{\zeta_2(z) - \zeta_1(z)} \sum_{N=0}^{\infty}\frac{\zeta^N}{\zeta_2^N(z)}  \Bigg)
\,, 
\end{align}
where
\begin{align}
\begin{split}
\zeta_1(z) &= - \frac{e^{-z t^2} \cosh(2 z t) + \Big(e^{2 z t^2} + e^{-2 z t^2} \sinh^2(2 z t)\Big)^{1/2}}{\sinh(2 z t^2)} \,, \\
\zeta_2(z) &= \frac{\Big(e^{2 z t^2} + e^{-2 z t^2} \sinh^2(2 z t)\Big)^{1/2} - e^{-z t^2} \cosh(2 z t)}{\sinh(2 z t^2)} \,.
\end{split}
\end{align}
From eqs. \eqref{ju}, we find that ${\cal Z}_N$ in eq. \eqref{m}, for $N=2, 3, \ldots$, is given explicitly by
\begin{align}
\begin{split}
\label{n}
{\cal Z}_N &= \frac{{\cal Z}^+_2(z,s_1) + {\cal Z}^-_2(z,s_1)}{2 (\zeta_2(z) - \zeta_1(z))} \Big[ \zeta_2(z) \zeta_1^2(z) \zeta_1^{-N}(z) - \zeta_1(z) \zeta_2^2(z)  \zeta_2^{-N}(z) \Big] \\
&+ {\mathbf 1}_{N \geq 3} \frac{e^{2 z t} \sinh(z t^2) {\cal Z}^+_2(z,s_1) + e^{- 2 z t} \sinh(z t^2) {\cal Z}^-_2(z,s_1)}{2 (\zeta_2(z) - \zeta_1(z))} \\
& \times \Big[ \zeta_2(z) \zeta_1^3(z) \zeta_1^{-N}(z) - \zeta_1(z) \zeta_2^3(z)  \zeta_2^{-N}(z) \Big] \,,
\end{split}
\end{align}
where ${\mathbf 1}_{N \geq 3} = 1$ for $N \geq 3$ and is equal to zero for $N =2$.

We notice next that $\zeta_1(z) < 0$ for any value of $z$,  while $\zeta_2$ is strictly positive, and that their ratio obeys
\begin{align}
\frac{\zeta_2((z)}{|\zeta_1(z)|} < 1
\end{align}
for any value of $z$ (note that this ratio vanishes for $z = 0$ and $z = \infty$) and $t$, which permits to rewrite formally the expression \eqref{n} as
\begin{align}
\label{nn}
{\cal Z}_N &= - \frac{{\cal Z}^+_2(z,s_1) + {\cal Z}^-_2(z,s_1)}{2 (\zeta_2(z) - \zeta_1(z))} \zeta_1(z) \zeta_2^2(z)  \zeta_2^{-N}(z) \left[ 1 + (-1)^N \frac{|\zeta_1(z)|}{\zeta_2(z)}  \left(\frac{\zeta_2(z)}{|\zeta_1(z)|}\right)^N \right] \nonumber\\
&- {\mathbf 1}_{N \geq 3} \frac{e^{2 z t} \sinh(z t^2) {\cal Z}^+_2(z,s_1) + e^{- 2 z t} \sinh(z t^2) {\cal Z}^-_2(z,s_1)}{2 (\zeta_2(z) - \zeta_1(z))} \zeta_1(z) \zeta_2^3(z)  \zeta_2^{-N}(z) \nonumber\\ & \times \left[1 - (-1)^N   \frac{\zeta_1^2(z)}{\zeta_2^2(z)}  \left(\frac{\zeta_2(z)}{|\zeta_1(z)|}\right)^N\right] \,,
\end{align}
Therefore, for $N \geq 3$, we have
\begin{align}
\label{n3}
{\cal Z}_N &= \Psi(z) e^{- N \ln(\zeta_2(z))}\,,
\end{align}
with
\begin{align}
\begin{split}
\label{p}
\Psi(z) &= - \frac{\zeta_1(z) \zeta_2^2(z)}{2 (\zeta_2(z) - \zeta_1(z))} \Bigg({\cal Z}^+_2(z,s_1) + {\cal Z}^-_2(z,s_1) \\ &+ \Big(e^{2 z t} \sinh(z t^2) {\cal Z}^+_2(z,s_1) + e^{- 2 z t} \sinh(z t^2) {\cal Z}^-_2(z,s_1)\Big) \zeta_2(z)\Bigg) \,.
\end{split}
\end{align}
Note that ${\cal Z}_N = 1$ when $z = 0$, as it should.

Now we are in position to determine the desired $\Phi_N$ in  eq. \eqref{phiN}. Inserting eqs. \eqref{n3} and \eqref{p} into eq. \eqref{phiN} and averaging over $s_1$, we have for $N \gg 1$
\begin{align}
\Phi_N &= \frac{1}{4} \left(\Psi_1 + \Psi_4 \right)\exp\Big(- \left(\lambda \cosh^2(J')  -  \Lambda_1(\lambda)\right) N\Big) \nonumber\\&+ \frac{1}{4} \Psi_2 \exp\left( - \left(\lambda \, e^{2J'} \cosh^2(J')  -  \Lambda_2(\lambda)\right) N\right) \nonumber\\
&+ \frac{1}{4} \Psi_3 \exp\left(- \left(\lambda \, e^{-2 J'} \cosh^2(J') -  \Lambda_3(\lambda) \right) N\right) \,,
\end{align}
where
\begin{align}
\begin{split}
\Psi_1 &= \Psi\Big(z = \lambda \cosh^2(J'), s_1=1\Big) \,,  \\
\Psi_2 &= \Psi\Big(z = \lambda e^{2 J'}\cosh^2(J'), s_1=-1\Big) \,,  \\
\Psi_3 &= \Psi\Big(z = \lambda e^{-2 J'} \cosh^2(J'), s_1=1\Big) \,,  \\
\Psi_4 &= \Psi\Big(z = \lambda \cosh^2(J'), s_1=-1\Big)    \,,
\end{split}
\end{align}
and
\begin{align}
\begin{split}
\Lambda_1 &= - \ln\Big(\zeta_2(z = \lambda \cosh^2(J'), s_1=1)\Big) \,, \\
\Lambda_2 &= - \ln\Big(\zeta_2(z = \lambda e^{2 J'} \cosh^2(J'), s_1=-1)\Big) \,, \\
\Lambda_3 &= - \ln\Big(\zeta_2(z = \lambda e^{- 2 J'} \cosh^2(J'), s_1=1)\Big) \,.
\end{split}
\end{align}
Consequently, the moment-generating function $\Phi_N$ is completely defined.

Lastly, we consider the moments of (not necessarily integer) positive order $n$ of the sequence-dependent current. 
This is done by multiplying the moment-generating function $\Phi_N$ by $\lambda^{n-1}$ and integrating over $\lambda$,
\begin{equation}
\label{B30}
\Big \langle j_{N}^{n} \Big \rangle = \frac{1}{\Gamma(n)} \left(\frac{c_0}{2 \delta t}\right)^n \int^{\infty}_0 \lambda^{n-1} \,  \Phi_N \, d\lambda \,.
\end{equation} 
The above integral is conveniently performed by changing the integration variable $\lambda = x/N$ and then expanding the integrand in the inverse powers of $N$, which yields our asymptotic expansions in eqs. \eqref{momentsj} and \eqref{jn}.

\section{Ternary alphabet}
\label{B}

Consider a bit more complicated case of a ternary alphabet which contains the letters $A$, $B$ and voids $\varnothing$, which appear with probabilities $(1-p)/2$, $(1-p)/2$ and $p$, respectively. The probability $p$ can be therefore interpreted as the concentration of letters $\varnothing$. Letters $A$ and $B$, present at equal amounts, on average, have a concentration $1 - p$.

We suppose that the interaction energy of the letters
on any two neighboring sites $x$ and $x +1$, which are brought in contact with each other in the course  of a hernia displacement along the chain 
is defined by the following rules: \\
The voids $\varnothing$ are not interacting neither between themselves nor with letters $A$ and $B$. Once any two $A$s (or two $B$s) appear close to each other due to a hernia, their interaction energy is equal to $u_0$, as in the case of a binary alphabet. Similarly, the interaction energy of a pair of an $A$ and a $B$ is equal to $u_{AB}$, (which, in general, is not equal to $u_0$). Correspondingly, the spin variable $\sigma_x$ is now a random, three-state variable of the form 
 \begin{equation}
 \sigma_x = \begin{cases}
\, \, 1  & \,\,\,\, \text{letter $A$, \,\,\, prob $(1-p)/2$} \,, \\
\, \, 0  & \,\,\,\, \text{letter $\varnothing$, \,\,\,  prob $p$} \,, \\
-1 & \,\,\,\, \text{letter $B$, \,\,\, prob $(1-p)/2$} \,,
\end{cases}
\end{equation}
and the energy $U_{x}$ of the hernia at site $x$
can be written as 
\begin{equation}
\label{B1}
U_{x} = J \sigma_x \sigma_{x+1} + K \sigma^2_{x} \sigma^2_{x+1}  \,,
\end{equation}
where the "Ising" coupling constant $J$ and the "bi-quadratic exchange" constant $K$ are given by eq. \eqref{00}. 
One may notice that such a choice of the interaction energy corresponds (expectedly) to the interaction energy of the spin-$1$ Blume-Emery-Griffiths model with zero crystal-field couplings \cite{0}. 

Respectively, for $U_{x}$ as defined in eq. \eqref{B1}, the transition probabilities 
 from $x$ to either $x+1$ or to $x - 1$, analogous to the ones defined in eq. \eqref{3}, are given by
\begin{equation}
\begin{split}
\label{B3}
p_{x,x+1} &= Z_x^{-1}   \exp\left(\frac{\beta}{2}\left[U_{x} - U_{x+1}  \right]\right) \\&= Z_x^{-1}\exp\left(J' \sigma_{x+1} \left(\sigma_{x} - \sigma_{x+2}\right) + K' \sigma^2_{x+1} \left(\sigma^2_{x} - \sigma^2_{x+2}\right) \right) \,, \\
p_{x,x-1} &= Z_x^{-1} \exp\left(\frac{\beta}{2} \left[U_{x} - U_{x-1} \right] \right) \\
&= Z_{x}^{-1} \exp\left(J' \sigma_{x} \left(\sigma_{x+1} - \sigma_{x-1}\right) + K' \sigma^2_{x} \left(\sigma^2_{x+1} - \sigma^2_{x-1}\right) \right) \,, K' = \frac{\beta K}{2} \,,
\end{split}
\end{equation}
where the normalization $Z_x$ is to be appropriately chosen to ensure that $p_{x,x+1}  + p_{x,x-1} = 1$.
Simplifying the above expressions, we have, similarly to eqs. \eqref{33}, 
\begin{align}
\begin{split}
p_{x,x+1} &= \frac{\phi_{x-1}}{\phi_{x-1} + \phi_{x+1}} \\
p_{x,x-1} &= \frac{\phi_{x+1}}{\phi_{x-1} + \phi_{x+1}} \,,
\end{split}
\end{align}
where the variable  $\phi_x$ has a bit more complicated form than in eq. \eqref{phi} and reads
\begin{equation}
\phi_x = \exp\Big(J' \sigma_x \sigma_{x+1} +  K' \sigma^2_x \sigma^2_{x+1}\Big) \,.
\end{equation}
In order to determine $\langle \tau_N \rangle$ and $\langle \tau_N^2\rangle$ we have to calculate the correlation functions $\langle \phi_0 \phi_1\rangle$,  $\langle 1/\phi_0 \phi_1\rangle$ and $\langle 1/\phi^2_0 \phi^2_1\rangle$ (see eq. \eqref{tau12}). This can be done very straightforwardly to give
\begin{align}
\begin{split}
\Big \langle \phi_0 \phi_1 \Big \rangle & = p + (1-p)  \left[p + (1-p) e^{K'} \cosh(J')\right]^2 \,, \\
\left \langle \frac{1}{ \phi_0 \phi_1} \right \rangle & = p + (1-p)  \left[p + (1-p) e^{-K'} \cosh(J')\right]^2 \,, \\
\left \langle \frac{1}{ \phi^2_0 \phi^2_1} \right \rangle & = p + (1-p)  \left[p + (1-p) e^{-2 K'} \cosh(2 J')\right]^2 \,.
\end{split}
\end{align}
Recalling the definition of $R_{\tau}$ (see eq. \eqref{R}), we find our eq. \eqref{Rtaup}.

\section{Sequence-dependent diffusion coefficient and function $\Theta_N$}
\label{D}

We concentrate here on the calculation of the partition function in eq. \eqref{Theta}. 
Firstly, in a standard way, we introduce auxiliary spin variables $s_x = \sigma_x \sigma_{x+1} = \pm 1$, $x = 1, 2, \ldots , N-1$ and $s_N = \sigma_N \sigma_1 = \pm 1$. 
In terms of these variables
\begin{align}
\begin{split}
&\frac{1}{N} \sum_{x=1}^N \frac{1}{\phi_x \phi_{x+1}} =  \cosh^2(J') - \frac{\sinh(2J')}{N} \sum_{x=1}^N s_x + \frac{\sinh^2(J')}{N} \sum_{x=1}^N s_x s_{x+1} \,,\\
&\frac{1}{N} \sum_{x=1}^{N} \phi_x \phi_{x+1} = \cosh^2(J') + \frac{\sinh(2J')}{N} \sum_{x=1}^N s_x + \frac{\sinh^2(J')}{N} \sum_{x=1}^N s_x s_{x+1} \,, 
 \end{split}
\end{align}
with $s_{N+1} = s_1$.
Consequently, the function in eq. \eqref{Theta} can be formally rewritten as
\begin{align}
\begin{split}
\Theta_N &= \exp\Big(- \left(\lambda_1 + \lambda_2\Big) \cosh^2(J')\right) \\&\times \sum_{\{s_x\}} \exp\Bigg(-\frac{\left(\lambda_1+\lambda_2\right)}{N} \sinh^2(J') \sum_{x=1}^N s_x s_{x+1} - \frac{\left(\lambda_2 - \lambda_1\right)}{N} \sinh(2 J') \sum_{x=1}^N s_x\Bigg) \,.
\end{split}
\end{align}
One may readily identify the sum with the subscript $\{s_x\}$ as being the partition function of a one-dimensional Ising model with magnetic field, which can be readily solved via the transfer matrix method to give, for $N \gg 1$, the following expression
\begin{align}
\Theta_N = \exp\Big(- (\lambda_1+\lambda_2) \cosh^2(J') + N \ln(\Lambda/2)\Big) \,,
\end{align}  
where $\Lambda$ is the largest eigenvalue of the transfer matrix given explicitly by
\begin{equation}
\Lambda = \exp\left({\cal J}\right) \cosh\left(h\right) + \Big(\exp\left(-2 {\cal J}\right) +  \exp\left(2 {\cal J}\right) \sinh^2\left(h\right)\Big)^{1/2} \,, 
\end{equation}
with 
\begin{equation}
{\cal J} = - \frac{(\lambda_1+\lambda_2)}{N} \sinh^2(J') \,, \quad h = \frac{(\lambda_1 - \lambda_2)}{N} \sinh(2 J') \,.
\end{equation}
Further on, noticing  that in the limit  $N \to \infty$, 
\begin{align}
\begin{split}
 \exp\Big(N \ln(\Lambda/2)\Big) &=  1 + \Big(3 \lambda_1^2 - 10 \lambda_1 \lambda_2 + 3 \lambda_2^2+ \left(5 \lambda_1^2 - 6 \lambda_1 \lambda_2 + 5 \lambda_2^2\right) \cosh(2 J')\Big)\\
& \times \frac{\sinh^2(J')}{4 N} + O\left(1/N^2\right) \,,
 \end{split}
\end{align}
we find that the large-$N$ asymptotic form of the function $\Theta_N$ is given by
\begin{align}
\begin{split}
\Theta_{N} &=  \exp\Big(- (\lambda_1+\lambda_2) \cosh^2(J')\Big) \Bigg(1 + \Big(3 \lambda_1^2 - 10 \lambda_1 \lambda_2 + 3 \lambda_2^2 \\&+ \left(5 \lambda_1^2 - 6 \lambda_1 \lambda_2 + 5 \lambda_2^2\right) \cosh(2 J')\Big)
\frac{\sinh^2(J')}{4 N} + O\left(1/N^2\right)\Bigg) \,.
\end{split}
\end{align}
Inserting the latter expansion into eq. \eqref{momk} and performing integrations, we find our eq. \eqref{Dmom}.

\end{document}